\documentclass[aps,showpacs,amsmath,twocolumn,superscriptaddress,longbibliography]{revtex4-1}

\usepackage{amsmath}
\usepackage{amssymb}
\usepackage{graphicx}
\usepackage{bbm}
\usepackage{bbold}

\newcommand{\vareps}{\varepsilon}

\begin{document}

\title{Spin-orbit coupling, quantum dots, and qubits in monolayer transition metal dichalcogenides}

\author{Andor Korm\'anyos}
\thanks{e-mail: andor.kormanyos@uni-konstanz.de}
\affiliation{Department of Physics, University of Konstanz, D-78464 Konstanz, Germany}

\author{Viktor Z\'olyomi}
\affiliation{Department of Physics, Lancaster University, Lancaster LA1 4YB, United Kingdom}

\author{Neil D. Drummond}
\affiliation{Department of Physics, Lancaster University, Lancaster LA1 4YB, United Kingdom}

\author{Guido Burkard}
\affiliation{Department of Physics, University of Konstanz, D-78464 Konstanz, Germany}

\begin{abstract}
 We derive an effective Hamiltonian which describes the dynamics of electrons in the 
 conduction band of monolayer transition metal dichalcogenides (TMDC) in the presence of perpendicular 
 electric and magnetic fields. We discuss in detail both the intrinsic   
 and the Bychkov-Rashba spin-orbit coupling (SOC) induced by an  external electric field. 
 We point out interesting  differences in the spin-split conduction band between 
 different TMDC compounds.   An  important consequence of the strong 
 intrinsic SOC is an effective out-of-plane $g$-factor for the electrons which
 differs from the free-electron $g$-factor $g\simeq 2$.
 We identify a new term in the Hamiltonian of the Bychkov-Rashba SOC which
 does not exist in III-V semiconductors. Using first-principles 
 calculations, we give estimates of the various parameters appearing in the theory. 
 Finally, we consider quantum dots (QDs) formed in TMDC materials and derive
 an effective Hamiltonian which allows us to calculate the magnetic field dependence 
 of the bound states in the QDs. We find that all states are both valley and spin split,
 which suggests that these QDs could be used as valley-spin filters.
We explore the possibility of using spin and valley states in TMDCs as quantum
bits, and conclude that, due to the relatively strong intrinsic
spin-orbit splitting in the conduction band, the most realistic
option appears to be a combined spin-valley (Kramers) qubit at low magnetic fields.

\end{abstract}


\maketitle

\section{Introduction}

Monolayers of transition metal dichalcogenides\cite{nnanotech-review} (TMDCs) posses a number 
of remarkable electrical and optical properties, which makes them an attractive research platform. 
Their material composition can be described by the formula $\textnormal{MX}_2$, where 
$\textnormal{M}=\textnormal{Mo}$ or $\textnormal{W}$ and 
$\textnormal{X}= \textnormal{S}$ or $\textnormal{Se}$. 
They are atomically thin, two-dimensional materials, and in contrast to graphene\cite{graphene-review}, 
they have a finite direct optical band gap of $\approx 1.5-2 \, {\rm eV}$, which is in the 
visible frequency range\cite{heinz-1,splendiani}.
This has facilitated the theoretical\cite{yao} and experimental\cite{heinz-2,cui-1,cao,sallen,cui-2,xiaodong-2}  
study of the rich physics related to the coupling  of the spin  and  the valley degrees of freedom.

Very recently, there has also been a growing interest in the transport properties of these materials. 
Although contacting and gating monolayer TMDCs is not entirely straightforward experimentally, 
progress is being made in this respect\cite{appenzeller,zhou,morpurgo,javey,kawakami,yuan,jarillo-herrero}. 
Electric\cite{yuan} and magnetic field\cite{kis,ye} effects are also studied currently, both in monolayer and 
few-layer samples. In addition, a promising experimental work 
has recently appeared regarding  spin-physics in these materials, 
showing, e.g., a viable method for spin-injection from ferromagnetic contacts\cite{kawakami}.

The finite band gap in the TMDCs should also  make it  possible  to confine 
the charge carriers with external gates  
and therefore to create, e.g.,  quantum dots. Together with the above 
mentioned progress in contacting and gating  TMDCs, this raises the
exciting question of whether these materials could be  suitable platforms to
host qubits \cite{loss-divincenzo}. 
Our work is motivated by this question.

\begin{figure}[htb]
 \includegraphics[scale=0.3]{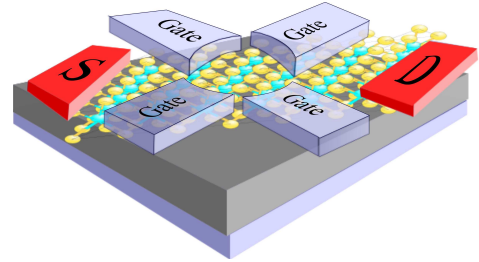}
 \caption{Schematics of a QD defined with the help of four top gates
   in a monolayer TMDC\@. S and D denotes the source and drain, respectively.}
 \label{fig:qmdot}
 \end{figure}
First, we are going to introduce an effective Hamiltonian which accurately describes the
physics in the conduction band  (CB) of TMDCs in the (degenerate) $K$ and $K'$ valleys 
of the Brillouin zone (BZ)\@.  
We confine our attention to the CB  while  the effect of the valence band (VB) and other relevant bands 
are taken into account through  an appropriate choice of the parameters appearing in the model. 
This approach is motivated by the facts that i)  the band-gap energy $E_{bg}$ is large with respect 
to other energy scales appearing in the problem, and ii) according to  
experimental observations,  the samples of TMDCs are often 
intrinsically $n$-doped\cite{kawakami,xiaodong-1} or show unipolar $n$-type behavior\cite{fuhrer}.
 To obtain realistic values of the parameters appearing in the theory
we have performed  density functional theory (DFT) calculations. 
We discuss  the important effects of the intrinsic SOC which manifest themselves both
through the spin-splitting of the CB and the different effective masses associated with the 
spin-split bands. We also point out that a perpendicular magnetic field, in addition
to the usual orbital effect, leads to the breaking of valley degeneracy. Moreover, due to 
the strong SOC,  the coupling of the spin degree of freedom to the magnetic field is described  by an 
out-of-plane  effective $g$-factor  $\tilde{g}^{\perp}_{sp}$.

We then study the effect of an external electric field and derive the Bychkov-Rashba SOC Hamiltonian
for TMDCs. This is motivated  by recent experiments\cite{xiaodong-1,xiaodong-2}, 
where strong electric fields were created by back gates to study the charged excitons.
In particular, we find that in contrast to III-V semiconductors and graphene, 
due to the lower symmetry of the system, 
the Bychkov-Rashba SOC Hamiltonian contains 
two terms, one of which has not yet been discussed in the literature. 
Using perturbation theory and first-principles (FP) calculations,
we can estimate the magnitude of this effect for each TMDC material. 

Finally, we consider QDs obtained by confining the charge carriers with 
gate electrodes.  We study the dependence of the spectrum of such
QDs on a perpendicularly applied external magnetic field.
We show that while pure spin and pure valley qubits are possible,
e.g., in small QDs in $\textnormal{MoS}_2$, but they require large
magnetic fields because of the relatively strong intrinsic SOC in the CB. 
On the other hand, combined spin-valley qubits
represented by a Kramers pair can be operated at small magnetic fields.
%
%Quantum dots
QDs in nanowires consisting of a $\textnormal{MoS}_2$ nanoribbon with
armchair edges or crystallographically aligned confining gates have been recently
discussed \cite{klinovaja}.    Our proposal  does not require 
atomically sharp boundaries or a precise control of the placement of the confining gates; 
therefore it should be easier to fabricate experimentally.
Moreover, we explicitly take into account the intrinsic spin-splitting of the CB.

The paper is organized as follows. In Sec.~\ref{sec:eff_Ham} we derive 
an effective Hamiltonian describing electrons in the
CB.  We take into 
account the effects of perpendicular external electric and magnetic fields.  
Using the results of FP  calculations we obtain values for 
the important parameters appearing in our model. 
In Sec.~\ref{sec:results} we use this model to study the magnetic field dependence
of the bound states in a QD\@. We also discuss the possible types of qubits that 
QDs in TMDCs can host. We conclude in Sec.~\ref{sec:summary}. 
In  Appendices \ref{sec:sevenband} and \ref{sec:eff-Ham-details} we present
the details of the derivation of the effective Hamiltonian. We collect some
useful formulas in Appendix \ref{sec:alpha-eigenfunc} and the details of our 
DFT calculations can be found in Appendix \ref{sec:comp-detail}.

%%%%%%%%%%%%%%%%%%%%%%%%%%%%%%%%%%%%%%%%%%%%%%%%%%%%%%%%%%%%%%%%%%%%%%%%%%%%%%%%%%%%%%%%%%%%%%%%%%%%%%%%%%%%%%%%%%%

\section{Effective Hamiltonian}
\label{sec:eff_Ham}

We  consider a monolayer TMDC and introduce a low-energy effective Hamiltonian which 
captures the most important effects in the spin-split conduction band  at the $K$ ($K'$) 
point. The detailed derivation of the model, which is based on a seven-band 
(without the spin degree of freedom) $\mathbf{k}\cdot\mathbf{p}$ Hamiltonian, 
is presented in Appendix \ref{sec:sevenband}. 
 \begin{figure}[htb]
 \includegraphics[scale=0.6]{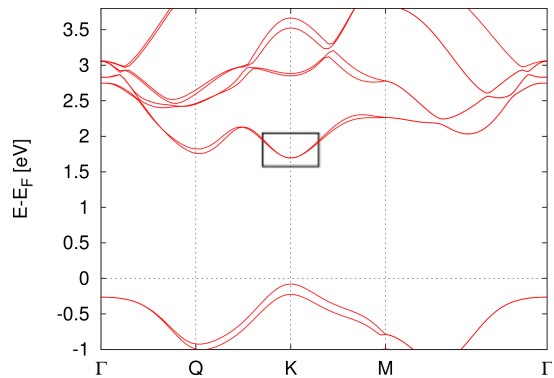}
 \caption{Spin-resolved band structure of $\textnormal{MoS}_2$ from DFT calculations. 
  The qualitative features of the band structure 
 are the same for all TMDCs. A blowup of the region in the black frame is shown in the upper panel of 
 Fig.~\ref{fig:CB-spin-split}.}
 \label{fig:mos2-bandstruct}
 \end{figure}
It is important to note that, as pointed out in Refs.~\onlinecite{song,our-mos2,cappelluti}, there are 
several  band extrema in the band structure of TMDCs  which can be of importance: 
see Fig.~\ref{fig:mos2-bandstruct}, where we show the band structure of $\textnormal{MoS}_2$ obtained
from DFT calculations.
Since we assume that the system is n-doped, the maximum at the $\Gamma$ point of the VB is 
not relevant. More important are the secondary minima in the CB, which are 
usually called the $Q$ (a.k.a.\ $T$) points. The exact alignment of the $Q$ point energy minimum with
respect to the $K$ point minimum is difficult to deduce from DFT and $GW$ calculations, because it depends 
quite sensitively on the details of these computations\cite{ashwin-1}. 
We have found that  using the local density approximation (LDA), all compounds, 
with the exception of  $\textnormal{MoS}_2$, 
become indirect gap semiconductors 
if we take into account the SOC, because the $Q$ point minimum is lower than the $K$ point minimum. 
More advanced $GW$ calculations also
give somewhat conflicting results and are quite sensitive to the level of 
theory\cite{compare-gw} ($G_0W_0$, $GW_0$ etc,) and the lattice constant used.  
Experimentally,  monolayer TMDCs 
show a significant increase of photoluminescence\cite{xiaodong-1,eda-2,terrones,cui-2} 
with respect to  few-layer or bulk TMDCs,
which is usually interpreted as  evidence that they are direct gap semiconductors. 
Therefore we assume that for low densities it is enough to consider only the $K$ and $K'$ points of 
the CB\@. 
For the formation of QDs from states around the $K$ point, the safest material appears to be
$\textnormal{MoS}_2$, where the secondary minima are most likely above the $K$ point minimum 
by a few hundred meV\cite{lambrecht,our-mos2}.  However, for operation
at low temperatures, the other TMDCs may also be suitable, as long as
the $Q$ point lies a few meV higher than the $K$ points.  In cases
where the $Q$ point lies below the $K$ point, one can envisage QDs
formed within the $Q$ valley, but this is beyond the scope of this paper.

%%%%%%%%%%%%%%%%%%%%%%%%%%%%%%%%%%%%%%%%%%%%%%%%%%%%%%%%%%%%%%%%%%%%%%%%%%%%%%%%%%%%%%%%%%%%%%%%%%%%%%%%%%%%%%%%%%%

\subsection{Electronic part and intrinsic spin-orbit coupling}
\label{subsec:h0_and_intr_so}

Due to the absence of a center of inversion and  strong SOC, the bands of monolayer TMDC materials 
are spin-split everywhere in the Brillouin zone (BZ), except at the high-symmetry 
points $\Gamma$ and  $M$, where the bands remain degenerate.
In addition,  the projection of the spin onto  the quantization axis perpendicular to the plane of the monolayer
is also preserved. This is a consequence of another symmetry, namely, the presence of a horizontal 
mirror plane $\sigma_h$. Therefore, a suitable basis to describe the CB is given by the 
eigenstates $\uparrow$, $\downarrow$ of the dimensionless spin Pauli matrix ${s}_z$ with eigenvalues $ s= \pm 1 $.
In what follows, we will often use the shorthand notation $\uparrow$ for $s=1$ and $\downarrow$ for $s=-1$.

In the absence of external magnetic and electric fields, the effective low-energy Hamiltonian which 
describes the spin-split CB  at the $K$ ($K'$) point in the basis $\uparrow, \downarrow$  is 
\begin{equation}
        \tilde{H}_{\rm el}^{\tau,s}+\tilde{H}_{\rm so}^{\rm intr}=        
	      \frac{\hbar^2 q_{+} q_{-}}{2 m_{\rm eff}^{\tau,s}} 
               + \tau \Delta_{cb} s_z   .
\label{kp-and-intr-so}            
\end{equation}
 Here, we introduce the inverse effective mass  
 $\frac{1}{m_{\rm eff}^{{\tau,s}}}=\frac{1}{m_{\rm eff}^0}-\tau s \frac{1}{\delta m_{\rm eff}}$, 
 where $\tau=1 (-1)$ for $K$ ($K'$) and  the wavenumbers $q_{\pm}=q_x \pm i q_{y}$ are measured 
 from the $K$ ($K'$) point. 
 Leaving the discussion of the effects of magnetic field to
 Sec.~\ref{subsec:magnetic_field}, we set
 $q_+ q_- = q_x^2+q_y^2$ and therefore the dispersion described by the Hamiltonian (\ref{kp-and-intr-so}) 
 is  parabolic and isotropic. The trigonal warping\cite{our-mos2}, which is much more pronounced in the 
 VB than in the CB, is neglected here. 
 
 The strong spin-orbit coupling in TMDCs has two consequences: firstly, as already mentioned, the CB 
 is spin-split at the $K$ ($K'$) point and this is described by the parameter $\Delta_{cb}$. 
 Secondly,  the effective mass is different 
 for the $\uparrow$ and $\downarrow$ bands.
 Our  sign convention for the effective mass assumes that the %$\uparrow$
 spin-up band is heavier than the %$\downarrow$ 
 spin-down band at the $K$ point (for details on the effective mass calculations see 
 Appendix \ref{sec:eff-Ham-details}).  
 The effective mass $m_{\rm eff}^{K,s}$ of different TMDCs, obtained from fitting the DFT band structure\cite{effmass-fit},
 is shown in Table \ref{tbl:effmass_and_so} (note that $m^{K',s}_{\rm eff}=m^{K,-s}_{\rm eff}$).
 As one can see, the difference between $m_{\rm eff}^{K,\uparrow}$ and  $m_{\rm eff}^{K,\downarrow}$
 is around $10-14\%$ for $\textnormal{MoS}_2$ and $\textnormal{MoSe}_2$, while it  is $\gtrsim 30\%$ 
 for the $\textnormal{WX}_2$ compounds.  
 In the seven-band $\mathbf{k}\cdot\mathbf{p}$ model this can be explained by  
 the fact that the effective mass  depends on the ratio of  the spin splittings in other bands 
(most importantly, in the VB and the second band above the CB) 
and the band gap $E_{bg}$. For the heavier compounds the spin-splittings are larger, but $E_{bg}$ remains 
roughly the same or even decreases, leading to a larger difference in the effective masses.
  \begin{table}[ht]
 \begin{tabular}{|c|c|c|c|c|}\hline
     & $\textnormal{MoS}_2$ & $\textnormal{WS}_2$  &  $\textnormal{MoSe}_2$ & $\textnormal{WSe}_2$ \\
  \hline
  $m_{\rm eff}^{K,\uparrow}/m_e$  & $0.49$ & $0.35$ & $0.64$ & $0.4$ \\
\hline
  $m_{\rm eff}^{K,\downarrow}/m_e$ & $0.44$ &  $0.27$ & $ 0.56$ & $0.3$ \\
  \hline
  $2 \Delta_{cb}$ [meV] & $3$ & $-38$ & $23$ & $-46$ \\
  \hline
\end{tabular}
\caption{Effective masses and  CB spin-splittings appearing in Hamiltonian (\ref{kp-and-intr-so}) for different 
TMDCs. $m_e$ is the free-electron mass.}
\label{tbl:effmass_and_so}
\end{table}

\begin{figure}[htb]
 \includegraphics[scale=0.45]{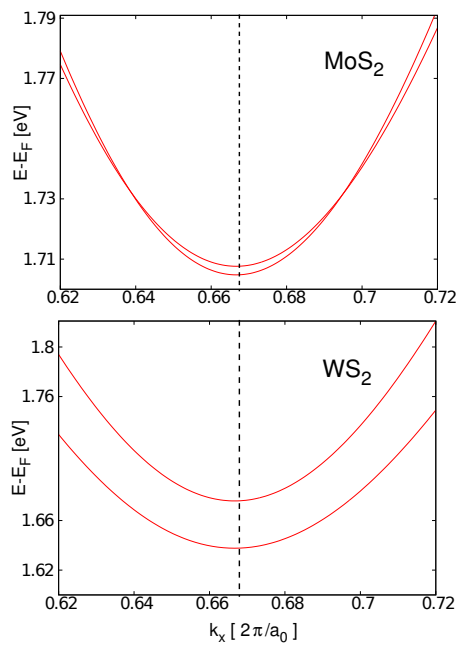}
 \caption{Upper panel: spin-split DFT CB of $\textnormal{MoS}_2$ in the vicinity of the $K$ point, which is 
 indicated by a vertical dashed line. Lower panel: the same  for $\textnormal{WS}_2$.  
 A band crossing, which  can be seen in the case of $\textnormal{MoS}_2$, is absent for 
 $\textnormal{WS}_2$.
 The small asymmetry in the figures with respect to the $K$ point, especially in the case of the
 band-crossing points in the upper panel, is  due to the fact that the  calculations 
 were performed along the $\Gamma K M$ line.}
 \label{fig:CB-spin-split}
 \end{figure}

 The results of DFT calculations also suggest that in the case of $\textnormal{MoX}_2$ materials there are
 band crossings between the spin-split CB  because the heavier band has higher 
 energy.  For   $\textnormal{WX}_2$ materials such a band crossing is absent. 
 Taking $\textnormal{MoS}_2$ and $\textnormal{WS}_2$ as an example, the dispersion in the vicinity
 of the $K$ point is shown in  Fig.~\ref{fig:CB-spin-split}. 
 A similar figure could be obtained for $\textnormal{MoSe}_2$ and $\textnormal{WSe}_2$
 as well, except that due to the larger spin splitting, the band crossings for $\textnormal{MoSe}_2$
occur further away from the $K$ point. Within the present model, which focuses on the CB, 
such a different behavior can be accounted for by a different sign of $\Delta_{cb}$
for   $\textnormal{MoX}_2$ and  $\textnormal{WX}_2$ materials. 
 A discussion about the possible microscopic origin of  this sign difference is presented 
 in  Appendix \ref{sec:eff-Ham-details}.  

 We note that a model Hamiltonian similar to Eq.~(\ref{kp-and-intr-so}), but without taking into
 account the difference in the effective masses, has  been used in Refs.~\onlinecite{wang,ochoa-2} to study
 spin-relaxation processes in  $\textnormal{MoS}_2$. 
 The effective mass difference and the sign of the effective SOC in the CB
 has also been discussed recently  in Ref.~\onlinecite{gui-bin}.

%%%%%%%%%%%%%%%%%%%%%%%%%%%%%%%%%%%%%%%%%%%%%%%%%%%%%%%%%%%%%%%%%%%%%%%%%%%%%%%%%%%%%%%%%%%%%%%%%%%%%%%%%%%%%%

\subsection{Effects of a perpendicular magnetic field}
\label{subsec:magnetic_field}

We assume that a homogeneous, perpendicular magnetic field of strength $B_z$ is applied. 
The $\mathbf{k}\cdot\mathbf{p}$ Hamiltonian can be obtained by using the
Kohn-Luttinger prescription, which amounts to replacing the numbers $q_x$ and $q_y$ in the above formulas
with operators: $\mathbf{q}\rightarrow \hat{\mathbf{q}}=\frac{1}{i}\boldsymbol{\nabla}+\frac{e}{\hbar}\mathbf{A}$, 
where $\mathbf{A}$ is the vector potential in Landau gauge and   $e>0$ is the magnitude of the electron charge.
Note that due to this replacement 
$\hat{q}_+$ and  $\hat{q}_-$ become non-commuting operators: 
$
[\hat{q}_-,\hat{q}_+] =\frac{2 e B_z }{\hbar},
$
where $|B_z|$ is the strength of the magnetic field. 
Therefore their order has to be preserved  when one folds down a multi-band Hamiltonian, which 
lies behind the  low-energy effective Hamiltonian (\ref{kp-and-intr-so}). 
As a consequence, for finite magnetic field further terms appear in the effective Hamiltonian.
The derivation of these terms within a seven-band $\mathbf{k}\cdot\mathbf{p}$ model 
is given in Appendix \ref{sec:eff-Ham-details}.

One finds that in an external magnetic field $H_{\rm el}^{\tau, s}$ in Eq.~(\ref{kp-and-intr-so}) is replaced 
by 
\begin{eqnarray}
 \tilde{H}_{\rm el}^{\tau, s}+\tilde{H}_{vl}^{\tau}+\tilde{H}_{sp}^{s} &=&
 \frac{\hbar^2 \hat{q}_{+}\hat{q}_{-}}{2 m^{\tau,s}_{\rm eff}} 
 + \frac{1+\tau}{2}\textnormal{sgn}(B_z)\hbar\omega_c^{\tau,s}\nonumber\\
&-&\frac{\tau}{2} \tilde{g}_{vl}\mu_{B}  B_z +  \frac{1}{2}\mu_B {g}_{so}^{\perp} s_z B_z 
\end{eqnarray}
where $\hbar \omega_c^{\tau,s}= e |B_z| /m_{\rm eff}^{\tau,s}$.

The term $\sim \omega_c^{\tau,s} $ in the bulk case  introduces a shift in the index of the 
Landau levels, so that  
there is an ``unpaired'' lowest Landau level in one of the valleys.  The next term,    
$\tilde{H}_{vl}^{\tau}=-\tau \tilde{g}_{vl}^{}\mu_{B} B_z$, breaks the valley symmetry of Landau levels. 
Here  
$\tilde{g}_{vl}^{}$ is the ``valley $g$-factor''. 
Similar effects have also been found in gapped monolayer\cite{koshino} and bilayer\cite{guido,arovas} graphene, and 
has  recently been noted for $\textnormal{MoS}_2$ as well\cite{asgari,rose,cai}; therefore we do not discuss them 
here in detail.  

A  new term, to our knowledge not yet considered in the literature of monolayer TMDC, is due to the strong SOC in 
these materials. It can be written in terms of an out-of-plane effective spin $g$-factor $g_{so}^{\perp}$:
$
\tilde{H}_{sp}^{s}= \frac{1}{2} \, {g}_{so}^{\perp}\, \mu_{B}\, s_z B_z
$
where  $\mu_{B}$ is the Bohr magneton.    
In addition, the well-known Zeeman term $H_Z= \frac{1}{2} \,g_e\, \mu_B\, s_z B_z\, $  also has to be taken 
into account\cite{dresselhaus-book}. Here $g_e \approx 2$ is the free-electron $g$-factor. 
The coupling of the spin to the magnetic field 
can therefore be described by 
%$
\begin{equation}
\tilde{H}_{sp, tot}^{s}=  \frac{1}{2}\tilde{g}_{sp}^{\perp}\mu_B s_z B_z,
\end{equation}
%$
where the total 
$g$-factor in the CB is $\tilde{g}_{sp}^{\perp}=g_e+g_{so}^{\perp}$. Values of   $\tilde{g}_{vl}^{}$
and $|{g}_{so}^{\perp}|$ obtained with the help of our DFT calculations are shown Table \ref{tbl:eff-g-factors}. 
\begin{table}[ht]
\begin{tabular}{|c|c|c|c|c|}\hline
     & $\textnormal{MoS}_2$ & $\textnormal{WS}_2$  &  $\textnormal{MoSe}_2$ & $\textnormal{WSe}_2$ \\
  \hline
   $\tilde{g}_{vl}^{}$ & $3.57$ &  $4.96$ & $3.03$ & $4.34$ \\
   \hline
   $|{g}_{so}^{\perp}|$  & $0.21$ & $0.84$ & $0.29$ & $0.87$ \\
  \hline
  $g_{vl}$ & $0.75$ & $1.6$ & $0.42$ &  $1.46$\\
  \hline
  $g_{sp}^{\perp}$  & $1.98$ & $1.99$ & $2.07$ & $2.04$  \\
  \hline
\end{tabular}
\caption{Valley  ($\tilde{g}_{vl}$, $g_{vl}$) and spin  ($g_{so}^{\perp}$, $g_{sp}^{\perp}$) 
$g$-factors for different TMDCs.}
\label{tbl:eff-g-factors}
\end{table}
The sign of ${g}_{so}^{\perp}$ cannot be obtained with our methods; it should be deduced either 
from experiments or from more advanced FP calculations. For the numerical calculations 
in Sec.~\ref{sec:qdot} we will assume that ${g}_{so}^{\perp}>0$.

In Sec.~\ref{sec:qdot} we will study the interplay of the magnetic field and the quantization due to 
confinement in QDs. 
While Eq.~(\ref{eq:CB-magnetic-Ham}) is a convenient  starting point to understand the Landau level physics, 
for relatively weak magnetic fields, when the effect of the confinement potential is important with respect to
orbital effects due to the magnetic field, one may 
re-write $\tilde{H}_{\rm el}^{\tau, s}$, $\tilde{H}_{vl}^{\tau}$, and $\tilde{H}_{sp,tot}^{s}$ in a 
slightly different form: 
\begin{eqnarray}
{H}_{\rm el}^{\tau, s}+{H}_{vl}^{\tau}+{H}_{sp,tot}^{s} &=&
 \frac{\hbar^2 \hat{q}_{+}\hat{q}_{-}}{2 m^{\tau,s}_{\rm eff}} 
 + \frac{1}{2}\textnormal{sgn}(B_z)\hbar\omega_c^{\tau,s}\nonumber\\
&+&\frac{\tau}{2} {g}_{vl}\mu_{B}  B_z +  \frac{1}{2}\mu_B {g}_{sp}^{\perp} s_z B_z,
\label{eq:CB-magnetic-Ham}
\end{eqnarray}
where ${g}_{vl}=(2 m_e/m_{\rm eff}^0)-\tilde{g}_{vl}$ and 
${g}_{sp}^{\perp}=\tilde{g}_{sp}^{\perp}-(2 m_e/\delta m_{\rm eff})$. This form 
shows explicitly that in contrast to ${H}_{\rm el}^{\tau, s}$, which 
depends on the product of $\tau$ and $s$ (through $m^{\tau,s}_{\rm eff}$), 
${H}_{vl}^{\tau}$  and ${H}_{sp,tot}^{s}$ depend only on $\tau$ and $s_z$, respectively.
This can help to understand the  level splittings patterns in QDs: see Sec.~\ref{sec:qdot}.
In particular, for  states which form a Kramers pair $\tau\cdot s=1$ or $-1$, therefore $H_{\rm el}^{\tau, s}$, 
which only depends on the product of $\tau$ and $s$, would not lift their degeneracy in the presence of
a magnetic field. 
Due to  $\tilde{H}_{vl}^{\tau}$, however, the degeneracy of the Kramers pair states 
will be lifted. Assuming $g_{so}^{\perp}>0$ and $B_z>0$, as in the calculations that lead to 
Figs.~\ref{fig:qm-mos2-40nm} and   \ref{fig:qm-ws2-40nm}, the values of $g_{vl}$ and $g_{sp}^{\perp}$ are shown 
in Table \ref{tbl:eff-g-factors}.

%%%%%%%%%%%%%%%%%%%%%%%%%%%%%%%%%%%%%%%%%%%%%%%%%%%%%%%%%%%%%%%%%%%%%%%%%%%%%%%%%%%%%%%%%%%%%%%%%%%%%%%%%%%%%% 
 
\subsection{External electric field and the Bychkov-Rashba SOC}
\label{subsec:ext_el_field}

The effective Hamiltonian (\ref{kp-and-intr-so}), describing the dispersion and the spin splitting of the 
CB is diagonal in spin space. 
An external electric field  has two effects: i) it can induce Bychkov-Rashba type  
SOC which will couple the different spin states, and ii)  it can change the energy of the band edge.  
We start with the discussion of the Bychkov-Rashba SOC\@.

For simplicity, we assume that the external electric field is homogeneous and that its strength 
is given by $E_z$.  Then the Bychkov-Rashba SOC in TMDCs is described by the Hamiltonian 
\begin{eqnarray}
 \tilde{H}_{\rm BR}^{\tau}&=&  
 \lambda_{\rm BR}^{i}\left(s_y q_x -s_x q_y\right) +
 \lambda_{\rm BR}^{r} \left(s_x q_x+ s_y q_y\right)
 \nonumber \\
 &=&
 \left(
 \begin{array}{cc}
  0 & \lambda_{BR}^{*}\, {q}_{-}^{}\\
  \lambda_{BR}^{}\, {q}_{+}^{} & 0
 \end{array}
 \right). 
 \label{bychkov-rashba-tmdc}
\end{eqnarray}
The first term, $\lambda_{\rm BR}^{i}\left(s_y q_x -s_x q_y\right)$, is the well-known Bychkov-Rashba\cite{bychkov-rashba1,*bychkov-rashba2} 
Hamiltonian, which is also present in GaAs and other III-V semiconductor compounds.  
It is equivalent to the  Bychkov-Rashba Hamiltonian
recently discussed in Ref.~\onlinecite{ochoa} in the framework of an effective  two-band model, 
which includes the VB\@. 
The second term, $\lambda_{\rm BR}^{r} \left(s_x q_x + s_y q_y\right)$, is also allowed by symmetry 
(see Table I of Ref.~\onlinecite{szunyogh}) because the 
pertinent symmetry group at the $K$ point in the presence of an external electric field is $C_3$. 
A derivation of  the Hamiltonian (\ref{bychkov-rashba-tmdc})  is given 
in Appendices \ref{sec:sevenband} and  \ref{sec:eff-Ham-details}. We note that 
the coupling constants $\lambda_{\rm BR}^{r}$ and  $\lambda_{\rm BR}^{i}$ cannot be tuned 
independently, because both of them are proportional to the electric field but with different 
proportionality factors. Using our microscopic model and FP calculations similar to those in
Ref.~\onlinecite{neil},  we can  estimate 
the magnitude of $\lambda_{\rm BR}$ but not $\lambda_{BR}^{r}$ and $\lambda_{BR}^{i}$ separately. 
The $|\lambda_{\rm BR}|$ values that we have obtained are shown in Table  \ref{tbl:lambda_br-values}. 
They give an upper limit for  the real values because we have neglected, e.g.,  screening in these 
calculations (for details see Appendix \ref{sec:eff-Ham-details}). 
More advanced DFT calculations, such as those recently done for bilayer graphene\cite{fabian_bilayer_SO},
would  be certainly of interest here. 
\begin{table}[htb]
 \begin{tabular}{|c|c|c|c|c|}\hline
     & $\textnormal{MoS}_2$ & $\textnormal{WS}_2$  &  $\textnormal{MoSe}_2$ & $\textnormal{WSe}_2$ \\
  \hline
  $|\lambda_{BR}|$ [eV\AA] & $0.033\, E_z$ & $0.13\, E_z$ & $0.055\, E_z$ & $ 0.18 \, E_z$ \\
\hline
\end{tabular}
\caption{Estimates of the Bychkov-Rashba SOC parameters  $|\lambda_{BR}|$. The perpendicular electric field
$E_z$ is in units of V/\AA.}
\label{tbl:lambda_br-values}
\end{table}

Comparing the numbers  shown in Table \ref{tbl:lambda_br-values} to the values found in 
InAs\cite{takayanagi} or InSb\cite{gilbertson},  
one can see that for relatively small values of the electric field ($E_z\lesssim 10^{-2}$ V/\AA), 
where the perturbation theory approach can be expected to work, $|\lambda_{BR}|$ is smaller by
an order of magnitude than in these semiconductor quantum wells.
Nevertheless, the Bychkov-Rashba SOC is important because it constitutes  an intra-valley 
spin-relaxation  channel, which does not require the simultaneous flip of spin 
\emph{and} valley. Thus, it may play a role in the quantitative understanding of the relaxation processes 
in the recent experiment of Jones \emph{et al.}\cite{xiaodong-2},
where a large back gate voltage was used. 

The external electric field has a further effect, which, however, turns out to be 
less important for our purposes. 
Namely, it shifts up the band edge of the CB, and the 
shift is, in principle, spin dependent [see Eqs.~(\ref{cb-U-K}), (\ref{cb-U-Kp}) 
in Appendix \ref{sec:eff-Ham-details}].  
The shift of the CB edge can be understood  in terms of  the  electric 
field dependence of the band gap (we note  that the band edge of the VB also depends
on the electric field, and the shifts of the VB and CB edges together would
describe the change of the band gap).  
In contrast to Ref.~\onlinecite{asgari}, however, in our model the shift of the band edge depends
quadratically on  the strength of the  electric field and not linearly. 
We think this is due to the fact that in the  model used in Ref.~\onlinecite{asgari} the $p$ orbitals of the
sulfur atoms are admixed only to the CB\@. In fact, symmetry considerations\cite{ochoa,our-mos2} and 
%our DFT calculations show that there is a small $p$ orbitals  weight of the $X$ atoms both in the VB and the CB\@.  
our DFT calculations show that the $p$ (or $d$) orbitals of the X atoms  have a small weight at the $K$ point 
\emph{both in the VB  and in the CB}\@.
Taking this into account, as in the tight-binding model of  Ref.~\onlinecite{cappelluti}, one would find that 
for weak electric field regime the dependence of the band gap is quadratic in the electric field. 
Moreover, both our perturbation theory and preliminary DFT results suggest that the shift of the band edge in the CB is 
actually very small, at least in the regime where the perturbation theory approach is applicable 
(see Appendix \ref{sec:eff-Ham-details} for details). Therefore we neglect it in the rest of the paper.  
The spin-dependence of the band-edge shift, being a higher-order effect, is expected to be even smaller.

%%%%%%%%%%%%%%%%%%%%%%%%%%%%%%%%%%%%%%%%%%%%%%%%%%%%%%%%%%%%%%%%%%%%%%%%%%%%%%%%%%%%%%%%%%%%%%%%%%%%%%%%%%%%%%%%%%%

\section{Results \label{sec:results}}

\subsection{Quantum dots in TMDCs}
\label{sec:qdot}

QDs  in novel low-dimensional structures, such as  bilayer graphene\cite{pereira,guido,vandersypen,yacoby}
and semiconductor nanowires with strong SOC\cite{nadj-perge1,*nadj-perge2,intronati}, are actively studied and the 
applicability of these structures for hosting  qubits has also been discussed. 
Motivated by the interesting physics revealed in these studies,  
we now consider QDs in two-dimensional semiconducting TMDCs defined by external electrostatic gates. 
In particular, we will be interested in the magnetic field dependence of the spectrum and discuss 
which eigenstates can be used as two-level systems for qubits.
We consider relatively small QDs which can be treated in the ballistic limit.  
The opposite limit, where disorder effects become important and the spectrum acquires certain 
universal characteristics,  can be treated along the lines of Ref.~\onlinecite{aleiner},  
but this  is beyond the scope of the present work. 

Nevertheless, based on  the findings of Sec.~\ref{subsec:h0_and_intr_so}, the following 
general considerations can be made: assuming a chaotic QD with mean level spacing 
 $\delta=2 \pi\hbar^2/(m_{\rm eff} A)$, where $A$ is the area of the dot, one can see that 
 one needs relatively small QD in order to make  $\delta$ larger than the 
thermal energy $k_B T$. For instance, taking a 
dot of radius $R\approx 40 \, {\rm nm}$ we find for, e.g.,
$\textnormal{MoS}_2$ that $\delta\approx 0.2 \, {\rm meV}$,  
corresponding to $T=2.3\, {\rm K}$, whereas for $\textnormal{WS}_2$, due to its smaller effective mass,
the mean level spacing is $T \approx 3.4 \, {\rm K}$. In this respect
TMDCs with smaller $m_{\rm eff}$, such as 
 $\textnormal{WS}_2$ and  $\textnormal{WSe}_2$, might be more advantageous. 
Although the required temperatures are smaller  than in the case of GaAs 
(which has $m_{\rm eff}\approx 0.067 m_e$), they are still achievable
with present-day techniques. 

In the following, for simplicity, we will study circular QDs because 
their spectrum can be obtained relatively easily and can illustrate some 
important features of  the spectrum of more general  cases.  
In particular, we will consider QDs in $\textnormal{MoS}_2$ and $\textnormal{WS}_2$.
The total Hamiltonian in the $K$, $K'$ valleys ($\tau=\pm 1$) reads  
\begin{equation}
 {H}={H}_{el}^{\tau,s}+\tilde{H}_{\rm so}^{\rm intr}+\tilde{H}_{BR}^{\tau}+{H}_{vl}^{\tau}
 +{H}_{sp,tot}^{}+V_{dot}
\end{equation}
where $V_{dot}$ is the confinement potential for the QD\@.
As we have shown, $\tilde{H}_{BR}^{\tau}$ is relatively small; therefore  we
 treat it as a perturbation,
whereas the stronger intrinsic SOI is treated exactly.
The Hamiltonian of the non-perturbed system is given by 
\begin{equation}
 {H}_{dot}={H}_{el}^{\tau,s}+{H}_{\rm so}^{\rm intr}
          +{H}_{vl}^{\tau}+{H}_{sp,tot}^{}+V_{dot},
\end{equation}
i.e., it is diagonal both in valley and in spin space. 
We consider a circular QD with hard wall boundary conditions: 
$V_{dot}(r)=0$ for $ r\leq R_d$ and $V_{dot}(r)=\infty$ if $r>R_d$. 
In cylindrical coordinates, the perpendicular magnetic field can be taken into account 
using  the axial gauge, where  $A_{\phi}=B_z r /2$ and $A_{r}=0$. With this choice,
since the rotational symmetry around the $z$ axis is preserved, ${H}_{dot}$ commutes
with the angular momentum operator $\hat{l}_z$ and they have common eigenfunctions.
The Schr\"odinger equation which determines the bound state energies and eigenfunctions 
can be solved by making use of the fact that, as 
noted in Ref.~\onlinecite{gogolin},  the operator $\hat{q}_+$  ($\hat{q}_-$)
appearing in  $H_{el}^{\tau}$ acts as a raising (lowering) operator on a suitably chosen trial function. 
Introducing the dimensionless new variable: 
$
\rho=\frac{1}{2} \left(\frac{r}{l_B}\right)^2
$, where $l_B=\sqrt{\frac{\hbar}{e B_z}}$ is the magnetic length, one finds for $B_z > 0$ that
\begin{subequations}
\begin{eqnarray}
 \hat{q}_{-}&=&\frac{-i}{l_B}\sqrt{\frac{\rho}{2}}e^{-i\varphi}
 \left(1+2 \partial_{\rho}-\frac{i}{\rho}\partial_{\varphi}\right)=\frac{-i \sqrt{2}}{l_B}\hat{\alpha}_{-}, \\
  \hat{q}_{+}&=&\frac{i}{l_B}\sqrt{\frac{\rho}{2}}e^{i\varphi}
 \left(1-2 \partial_{\rho}-\frac{i}{\rho}\partial_{\varphi}\right)=\frac{i \sqrt{2}}{l_B}\hat{\alpha}_{+}.
\end{eqnarray}
\label{alpha-opr} 
\end{subequations}
The eigenfunctions of the operators 
$\hat{\alpha}_{+}$ and  $\hat{\alpha}_{-}$, which are (i) regular at $\rho=0$ and (ii) also eigenfunctions 
of $\hat{l}_z$,  are
$
g_{a,l}(\rho,\varphi)=e^{i l \varphi} \rho^{\frac{|l|}{2}} e^{-\frac{\rho}{2}} M(a,|l|+1,\rho)
$, 
where $l$ is an integer  and 
$M(a,|l|+1,\rho)$ is the confluent hypergeometric function  of the first 
kind\cite{confluent}. One can show that 
\begin{equation}
 \hat{\alpha}_{+}\hat{\alpha}_{-}\,g_{a,l}(\rho,\varphi) = 
 \left\{
 \begin{array}{cc}
  -a\, g_{a,l}(\rho,\varphi)  & \mbox{if~} l\le 0 \\
   (l-a)\, g_{a,l}(\rho,\varphi)  & \mbox{if~} l>0.
 \end{array}
 \right.
\end{equation}
(For details see Appendix \ref{sec:alpha-eigenfunc}.)
Considering now the Schr\"odinger equation for the bulk problem, i.e., for $V_{dot}=0$ 
in valley $\tau$ for spin $s$, it reads  
\begin{eqnarray}
 [\hbar\omega_{c}^{\tau,s}\hat{\alpha}_{+}\hat{\alpha}_{-} &+& \frac{1}{2} {\rm sgn}(B_z) \hbar\omega_c^{\tau,s} +
 \tau\Delta_{cb} s_z \nonumber \\ 
 &+&(\frac{\tau}{2} g_{vl}^{}\mu_{vl} + \frac{1}{2} g_{sp}^{\perp} \mu_B s_z) B_z]\Psi= E \Psi,
\end{eqnarray}
where $\Theta(x)$ is the Heaviside step function. 
The wave functions
$
\Psi^{\uparrow}_{l}(\rho,\varphi)= \frac{e^{i l \varphi}}{\sqrt{2\pi}}
\left(
\begin{array}{c}
1\\
0
\end{array}
\right) \Phi^{}_{l}(\rho)
$ and 
$\Psi^{\downarrow}_{l}(\rho,\varphi)= \frac{e^{i l \varphi}}{\sqrt{2\pi}}
\left(
\begin{array}{c}
0\\
1
\end{array}
\right) \Phi^{}_{l}(\rho)
$
will be eigenfunctions if 
$\Phi^{}_{l}(\rho)=
\rho^{\frac{|l|}{2}} e^{-\frac{\rho}{2}} M(a^{}_{l},|l|+1,\rho)$
and 
\begin{equation}
\hbar\omega_c^{\tau,s} a^{}_{l}=
\left\{
\begin{array}{cc}
E^{\tau,s}_{} & \mbox{if}\qquad l \le 0\\
E^{\tau, s}_{} + l\hbar\omega_c^{\tau,s}  &  \mbox{if}\qquad l>0.
\end{array}
\right.
\label{E_tau_s}
\end{equation}
Here 
$
E^{\tau,s}_{}=
1/2 {\rm sgn}(B_z) \hbar \omega_c^{\tau,s}+ \tau\,s\, \Delta_{cb} + \frac{1}{2}(\tau\, g_{vl}^{}\mu_{vl}+ s\, g_{sp}^{\perp} \mu_B)B_z-E
$.
The bound state solutions of the QD problem are determined by the condition that the 
wave function has to vanish at $r=R_d$, i.e., one has to find the energy $E^{\tau,s}_{l}$ for which 
 $M(a^{}_{l},|l|+1,\rho[r=R_d])=0$. The task is therefore to find for a given magnetic field $B_z$  
 and quantum number $l$ the roots of $M(a^{}_{l},|l|+1,\rho[r=R_d])=0$ as a function of $a^{}_{l}$. 
 The $a^{}_{l}$ values can be calculated numerically. Once the $n$th root 
 $a^{}_{n,l}$ is known, the energy of the bound state $E_{n,l}^{\tau,s}$ can be expressed
 using Eq.~(\ref{E_tau_s}).
 
  \begin{figure}[t]
 \includegraphics[scale=0.5]{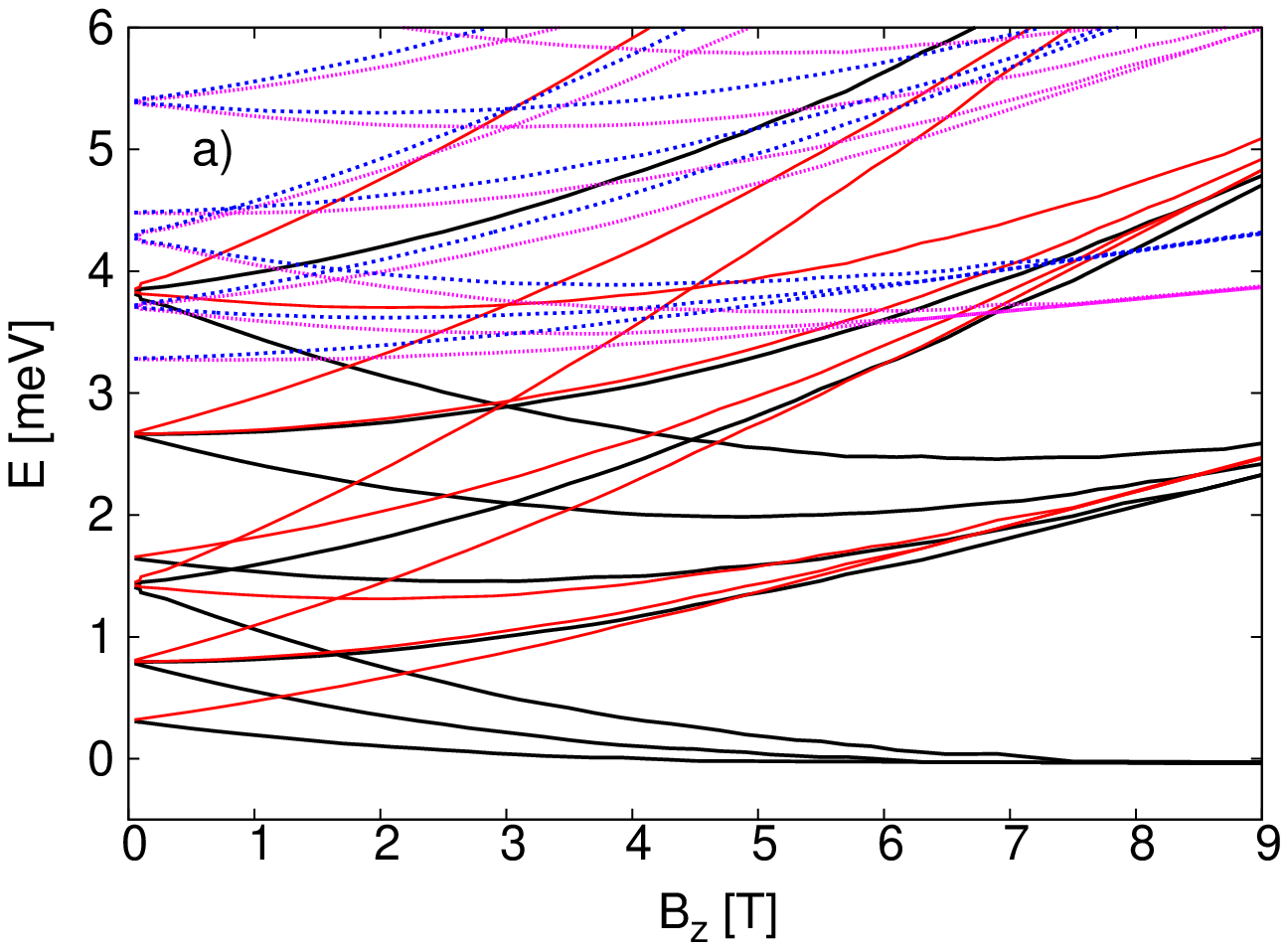}
 \includegraphics[scale=0.5]{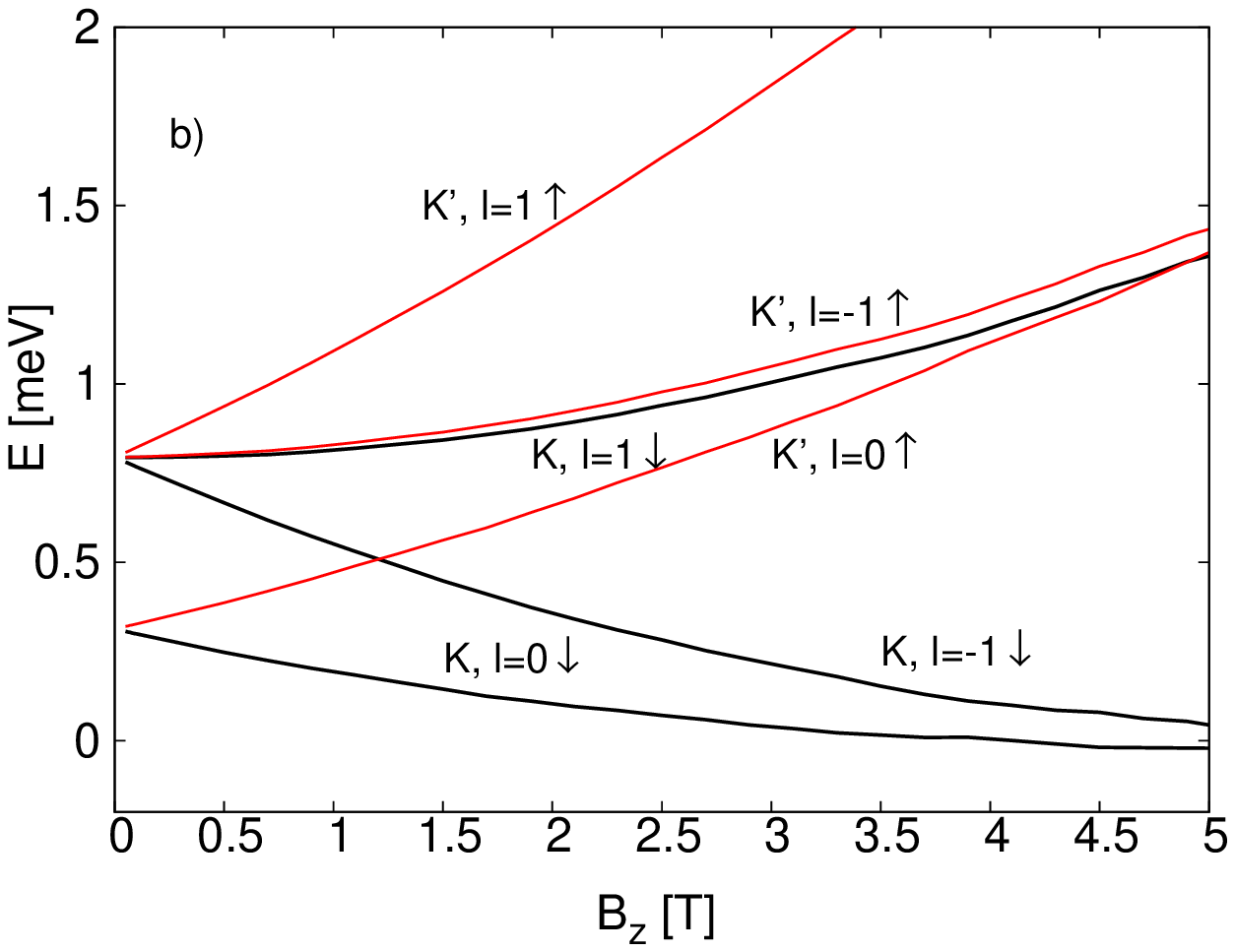}
 \caption{ (a) Spectrum of a $\textnormal{MoS}_2$ QD of radius $R_d=40 \, {\rm nm}$ as a function of the 
  perpendicular magnetic field $B_z>0$.  Black (purple) lines: spin $\downarrow$ ($\uparrow$) in the $K$ valley.  
  Red (blue) lines: spin $\uparrow$ ($\downarrow$) in the $K'$ valley.
  States up to $|l|=2$ and $n=2$ are shown. 
  (b) Part of the spectrum shown in (a) for small magnetic fields and low energies. Labels show the valley, 
  orbital quantum number $l$, and spin state for each level. 
  The values of $m_{\rm eff}^{\tau,s}$, $g_{vl}$ and $g_{sp}^{\perp}$ used in
  the calculations can be found in Tables \ref{tbl:effmass_and_so} and \ref{tbl:eff-g-factors}. 
}
 \label{fig:qm-mos2-40nm}
 \end{figure}

The numerically calculated spectrum for a QD with $R_d=40 \,{\rm nm}$ in 
$\textnormal{MoS}_2$ is shown in Fig.~\ref{fig:qm-mos2-40nm}(a). 
At zero magnetic field, because of the quadratic dispersion in our model, 
there is an effective time reversal symmetry acting within each valley and therefore  
states with angular momentum $\pm l$ within the same valley are degenerate.
For finite magnetic field all levels are both \emph{valley and spin} split.  
For even larger magnetic fields, when $l_B\lesssim R_d$, the dot levels merge into Landau levels. 
Since $\Delta_{cb}$ is relatively small with respect to the cyclotron energy $\hbar\omega_c^{\tau,s}$, 
spin-split states $\downarrow$ and $\uparrow$ from the same valley can cross at some larger, 
but still finite magnetic field (see, e.g., the crossing between the black and green lines for 
$E>3\, {\rm meV}$ for states in valley $K$ in Fig.~\ref{fig:qm-mos2-40nm}a). 

Taking into account the  Bychkov-Rashba SOC turns the crossings between 
states $|a, l,\uparrow \rangle$ and $|a, l+1,\downarrow \rangle$, $l\ge 0$ 
into avoided crossings. The selection rules for $H_{BR}^{\tau}$
can be derived by rewriting  $\tilde{H}_{BR}^{\tau}$ in terms of the operators 
$\alpha_{-}$ and $\alpha_{+}$ and calculating their effect on the non-perturbed eigenstates 
(see Appendix \ref{sec:alpha-eigenfunc} for details).  
For the low-lying energy states, in which we are primarily interested,
the effect of the Bychkov-Rashba SOC is to introduce  level repulsion  
between these states and higher energy ones allowed by the selection rules. 
Taking  $|\lambda_{BR}|/l_B$ as a characteristic energy scale of this coupling and using  
Table \ref{tbl:lambda_br-values}
one can see that for magnetic fields $\lesssim 10\,{\rm T}$ and electric fields 
$E_z\lesssim 10^{-2}\,{\rm V/\AA}$ the level repulsion is much smaller than the spin 
splitting $\Delta_{cb}$ and therefore we neglect it. 
 
Figure ~\ref{fig:qm-mos2-40nm}(b) shows the low-field and low-energy regime of Fig.~\ref{fig:qm-mos2-40nm}(a). 
As one can see, for $B_z\gtrsim 1 \, {\rm T}$ the  lowest energy states reside in valley $K$. 
We emphasize that,  in contrast to gapped monolayer\cite{schnez,guido,recher} and bilayer\cite{guido,recher} graphene,
the energy states are also spin polarized.  
This suggest that QDs in $\textnormal{MoS}_2$ can be used as   simultaneous valley and spin filters.

 \begin{figure}[t]
  \includegraphics[scale=0.5]{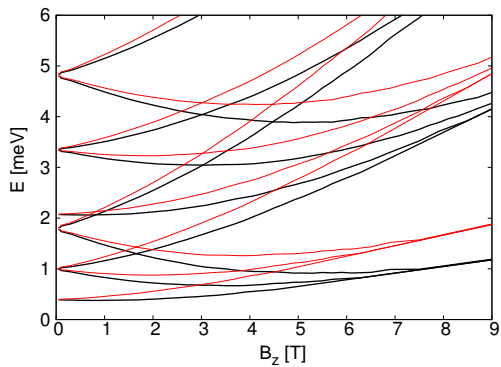}
  \caption{ Spectrum of a 40 nm $\textnormal{WS}_2$ QD as a function of the 
   perpendicular magnetic field $B_z>0$. Black (red) lines show the spin $\uparrow$  ($\downarrow$) 
   states from valley $K$ ($K'$). The values of $m_{\rm eff}^{\tau,s}$ can be found in 
   Table \ref{tbl:effmass_and_so}, whereas 
    $g_{vl}=1.6$, and $g_{sp}^{\perp}=1.99$ (see Table \ref{tbl:eff-g-factors}).
}
  \label{fig:qm-ws2-40nm}
  \end{figure}

Figure~\ref{fig:qm-ws2-40nm} shows the low-energy spectrum of a
$\textnormal{WS}_2$ QD with radius $R_d=40 \, {\rm nm}$.
Qualitatively, it is  similar to $\textnormal{MoS}_2$, but because the spin splitting $\Delta_{cb}$ 
between the $\uparrow$ and $\downarrow$ states belonging to the same valley is much larger than was 
the case for $\textnormal{MoS}_2$, 
they do not cross for the magnetic field range shown in Fig \ref{fig:qm-ws2-40nm}.  One can also observe 
that the $B_z=0$ level spacing is somewhat larger than in the  $\textnormal{MoS}_2$ QD
[see Fig.~\ref{fig:qm-mos2-40nm}(b)]. 
Another important observation that can be made by comparing the results for  $\textnormal{MoS}_2$ and 
$\textnormal{WS}_2$ is the following: for a given magnetic field,
e.g., $B_z=5 \, {\rm T}$, 
the splitting between  states belonging to different valleys is significantly larger for 
the former  material than for the latter 
(compare Figs.~\ref{fig:qm-mos2-40nm}(b) and ~\ref{fig:qm-ws2-40nm}). 
This is due to  the  different sign of $\Delta_{cb}$ and hence different spin polarization of the 
lowest levels in the two materials: in the case of  $\textnormal{MoS}_2$ the valley splitting 
(described by $H_{vl}^{\tau}$) and the coupling of the spin to the magnetic field (given by $H_{sp,tot}$) 
reinforce each other, whereas for $\textnormal{WS}_2$ they  counteract, and since 
$g_{vl}$ and $g_{sp}^{\perp}$ have similar magnitude, in the end the valley splitting of the levels at large magnetic 
fields is small. 
This  suggests that for spin and valley filtering  the  $\textnormal{MoX}_2$ compounds are better suited. 

The  qualitative difference  between $\textnormal{MoS}_2$ and $\textnormal{WS}_2$ regarding the valley splitting 
does not depend crucially on the exact
values of the bulk parameters $\tilde{g}_{vl}$ and $g_{so}^{\perp}$. 
However, on a more quantitative level, the valley splitting does depend on the exact values 
of the valley and spin $g$-factors, which were calculated  using the DFT band gap 
and the $\mathbf{k}\cdot\mathbf{p}$  parameter $\gamma_3$ 
(see Sec.~\ref{sec:eff-Ham-details} for details). It is known that DFT underestimates the band gap, 
and the value of $\gamma_3$  depends to some extent   on the way it is extracted from the 
FP computations. 
As a result, the values shown in Table~\ref{tbl:eff-g-factors}
probably overestimate  $\tilde{g}_{vl}$ and $g_{so}^{\perp}$. 
\begin{figure}[t]
  \includegraphics[scale=0.5]{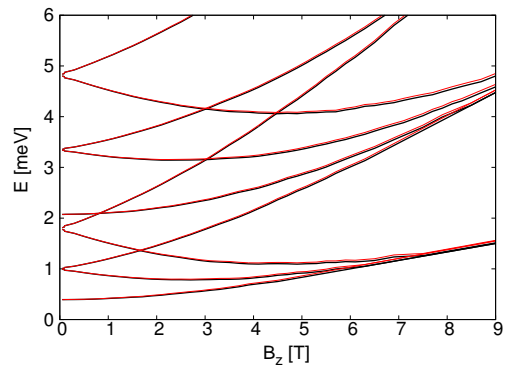}
  \caption{ Spectrum of a 40 nm $\textnormal{WS}_2$ QD as a function of the 
   perpendicular magnetic field $B_z>0$.  
   The values of $m_{\rm eff}^{\tau,s}$ can be found in Table \ref{tbl:effmass_and_so} and 
    we used ${g}_{vl}=2.31$ and $g_{sp}^{\perp}=1.84$ (c.f.  Fig.~\ref{fig:qm-ws2-40nm}).  
   Black (red) lines show spin $\uparrow$  ($\downarrow$) 
   states from the $K$ ($K'$) valley.}
  \label{fig:ws2-40nm-altern}
  \end{figure}
To illustrate this point, we show in Fig.~\ref{fig:ws2-40nm-altern} the low-energy spectrum of the same
$\textnormal{WS}_2$ quantum dot as in Fig.~\ref{fig:qm-ws2-40nm} 
but using a $g_{vl}$  ($g_{sp}^{\perp}$) which was obtained from a 
$\tilde{g}_{vl}$  ($g_{so}^{\perp}$) that is   $\sim 20\%$ smaller than the one shown in Table~\ref{tbl:eff-g-factors}.
The valley splitting of the bound states can now barely be observed.

\subsection{Qubits in TMDC quantum dots}

Circular hard-wall QDs in two-dimensional semiconducting TMDCs have a
spectrum similar to the characteristic Fock-Darwin spectrum for harmonically
confined QDs  (Fig.~\ref{fig:qm-mos2-40nm}).   Taking $\textnormal{MoS}_2$ as an example, 
due to the intrinsic spin-orbit splitting of about 3\,meV, each of the
spin- and valley-degenerate states $|l\rangle$ 
%is labeled by the angular momentum quantum number $l$ and 
splits into two Kramers pairs at vanishing
magnetic field $B=0$, namely ($|l,K,\uparrow\rangle$,
$|l,K',\downarrow\rangle$) and ($|l,K',\uparrow\rangle$,
$|l,K,\downarrow\rangle$).
Only at relatively high magnetic fields do
we observe a crossing of two states with the
same spin and opposite valley %(around $B\simeq 17\,{\rm T}$) 
or within the same valley with opposite spin. 
%(around $B\simeq 23\,{\rm T}$). 
These valley and spin pairs could serve as valley or spin
qubits, respectively, but the required high magnetic field and
the other overlapping levels with  different $l'$ quantum numbers complicate
their realization. (The energy of higher angular momentum states
can in principle be increased by making the QD smaller).

In view of the above, the  most realistic approach seems to be  to use the 
lowest Kramers pairs around $B=0$, e.g., $|l=0,K',\uparrow\rangle$ and
$|l=0,K,\downarrow\rangle$ as a combined spin-valley qubit
\cite{nadj-perge1,*nadj-perge2,flensberg}.  The energy splitting of these two-level systems
could be tuned using the external magnetic field.
The relaxation time of such spin-valley qubits in TMDC QDs will be
limited only by the longer spin or valley relaxation time, while the 
pure dephasing time will be limited by the shorter of the two.
The exchange interaction then provides the necessary coupling of adjacent spin-valley qubits for the
realization of two-qubit gates.

%%%%%%%%%%%%%%%%%%%%%%%%%%%%%%%%%%%%%%%%%%%%%%%%%%%%%%%%%%%%%%%%%%%%%%%%%%%%%%%%%%%%%%%%%%%%%%%%%%%%%%%%%%%%%%%%%%%
 
\section{Summary}
\label{sec:summary}

In summary, we have studied  TMDCs as possible host materials for QDs and qubits.
We considered n-doped samples, which can 
be described by an effective model which involves only the CB\@. 
Using our FP calculations, we have obtained the parameters 
that appear in  the effective Hamiltonian (effective masses, $g$-factors) 
for four distinct TMDC materials.
We discussed the effects of external magnetic and electric fields,
pointing out that the former leads to the splitting of the energy levels in different
valleys, while the latter induces a Bychkov-Rashba SOC, which, however, appears to 
be rather small.   
We have used the effective Hamiltonian to calculate the spectrum of 
circular QDs, finding  that all bound states are both spin and 
valley split. Our results  suggest that at large magnetic field QDs in TMDCs can
be used as spin and valley filters, but that this effect may depend on material-specific details. 
Finally, we have discussed the possible types of qubits that 
QDs in TMDC materials can host.  We have found that Kramers pairs around
$B_z=0$ appear to be the most realistic candidates.  

The effective one-band model and
the material parameters that we obtained for different TMDCs will hopefully be helpful 
in other fields as well, e.g., for studying  plasmonic excitations\cite{schliemann}.

\emph{Note added} After the submission of this work another manuscript appeared on the 
arXiv and has been subsequently published\cite{kosmider-2} on the spin-splitting 
in the conduction band of monolayer TMDCs.

\section{Acknowledgments}

We acknowledge discussions with Lin Wang. 
A.\ K.\ and G.\ B.\ acknowledge funding from  DFG under programs SFB767, SPP1285, FOR912 
and from the European Union through Marie Curie ITN ${\rm S}^3$NANO\@.
V. Z.  acknowledges  support from the Marie Curie project CARBOTRON\@.

\appendix

%%%%%%%%%%%%%%%%%%%%%%%%%%%%%%%%%%%%%%%%%%%%%%%%%%%%%%%%%%%%%%%%%%%%%%%%%%%%%%%%%%%%%%%%%%%%%%%%%%%%%%%%%%%%%%%%%%%

\section{Seven-band model}
\label{sec:sevenband}

\subsection{Introduction}
\label{subsec:sevenband-intro}

Our aim is to derive a low-energy effective Hamiltonian valid close to the $K$ ($K'$) point of the BZ,
which describes  the band dispersion, the effects of intrinsic SOC, and the SOC induced by 
an external electric field (Bychkov-Rashba effect). To this end 
we will consider the SOC in the atomic approximation, apply  $\mathbf{k}\cdot\mathbf{p}$ perturbation 
theory,  and take into account the effect of an external electric field perturbatively. 
We consider a seven-band model (without spin) 
which contains every band from the third band below the VB (which we call VB-3) up to the second band 
above the CB (denoted by CB+2 henceforth), i.e., we take the basis
 $\{ |\Psi_{E_2^{'}}^{vb-3},s\rangle, |\Psi_{E_1^{''}}^{vb-2},s\rangle, |\Psi_{E_2^{''}}^{vb-1},s\rangle, 
|\Psi_{A'}^{vb},s\rangle, |\Psi_{E_1^{'}}^{cb},s\rangle, \\|\Psi_{A^{''}}^{cb+1},s\rangle,
|\Psi_{E_1^{'}}^{cb+2},s\rangle
\}$. 
The upper index $b=\{vb-3,\\ vb-2,vb-1,vb,cb,cb+1,cb+2\}$ denotes the band and the lower 
index $\mu$ indicates the pertinent irreducible representation of the point group 
$C_{3h}$, which is the pertinent symmetry group for the unperturbed basis functions 
at the $K$ point of the BZ\@.
The spinful symmetry basis functions are represented by 
$| \Psi_{\mu}^{b}, s\rangle =  | \Psi_{\mu}^{b}\rangle \otimes | s \rangle$, 
where $s=\{ \uparrow,\downarrow \}$ denotes the spin degree of freedom. 
Note, that the basis states can be separated into two groups. The first group contains 
those states whose orbital part  is symmetric with respect  
to the mirror operation $\sigma_h$: 
$
\{ |\Psi_{A'}^{vb},s\rangle, |\Psi_{E_1^{'}}^{cb},s\rangle,
|\Psi_{E_2^{'}}^{vb-3},s\rangle,|\Psi_{E_1^{'}}^{cb+2},s\rangle
\}
$;  
 the second group contains  antisymmetric states: 
$
\{|\Psi_{E_1^{''}}^{vb-2},s\rangle, |\Psi_{E_2^{''}}^{vb-1},s\rangle,
|\Psi_{A^{''}}^{cb+1},s\rangle
\}
$.

%%%%%%%%%%%%%%%%%%%%%%%%%%%%%%%%%%%%%%%%%%%%%%%%%%%%%%%%%%%%%%%%%%%%%%%%%%%%%%%%%%%%%%%%%%%%%%%%%%%%%%%%%%%%%%%%%%%%%%%%%%%5

\subsection{Intrinsic spin-orbit coupling at the $K$ ($K'$) point of the Brillouin zone}
\label{subsec:SOC}

The intrinsic SOC is treated in the atomic approximation, whereby
the SOC is given by the Hamiltonian\cite{dresselhaus-book}
\begin{equation}
 \mathcal{H}_{\rm so}^{\rm at}=\frac{\hbar}{4 m_e^2 c^2} \frac{1}{r} \frac{d V(r)}{d r} \,
 \mathbf{\hat{L}} \cdotp \hat{\mathbf{S}}.
 \label{atomic-SOC}
\end{equation}
Here $V(r)$ is the spherically symmetric atomic potential, $\mathbf{\hat{L}}$ is the  angular momentum operator and 
$\mathbf{\hat{S}}=(s_x,s_y, s_z)$ is a vector of spin Pauli matrices $s_x,\,s_y,\, s_z$ (with eigenvalues $\pm 1$).
One can rewrite the product $\mathbf{\hat{L}} \cdotp \mathbf{\hat{S}}$ as 
$\mathbf{\hat{L}} \cdotp \mathbf{\hat{S}}=\hat{L}_z s_z + \hat{L}_{+} s_{-} + \hat{L}_{-} s_{+}$, where 
$\hat{L}_{\pm}= \hat{L}_{x} \pm i \hat{L}_{y}$ and 
$s_{\pm}=\frac{1}{2}(s_{x}\pm i s_{y})$. The task is then to calculate the matrix elements of (\ref{atomic-SOC}) 
in the  basis introduced in Sec.~\ref{subsec:sevenband-intro} at the  $K$ ( $K'$)  point of the BZ.
To this end one can make use of the symmetries of the band-edge wave functions. 
For instance, the diagonal matrix elements are proportional to $s_z$, this is because the $\hat{L}_z$ is symmetric
with respect to $\sigma_h$ whereas $\hat{L}_{\pm}$ is antisymmetric. Conversely, most of the off-diagonal
matrix elements will be proportional to $s_{\pm}$, reflecting the fact that they are related to matrix elements
having different symmetry with respect to $\sigma_h$. The only exception is the off-diagonal matrix element 
between $ |\Psi_{E_2^{'}}^{v-3},s\rangle$ and $|\Psi_{E_1^{'}}^{c+2},s\rangle$, which connects symmetric states. 
In addition, one has to consider the transformation properties of the basis functions and angular momentum operators 
with respect to a rotation by $2\pi/3$. 
The general result for the $K$ point is shown in Table \ref{tbl:SOC-at-K}.

Before showing  further details of the calculations in subsections 
\ref{subsec:kp-matrix} and \ref{subsec:efield},  some comments are in order here. 
As long as one considers states close to the $K$ point, the largest energy scale 
is the band gap and other band-edge energy differences. The next largest energy scale
comes from the SOC. As an upper limit of the various diagonal and off-diagonal 
matrix elements (see Table \ref{tbl:SOC-at-K}) one can take the spin-splitting of the 
VB.  The reason is that the main contribution to this band at the $K$ point 
comes from the metal $d$ orbitals and the metal atoms, being much 
heavier than the chalcogenides, are expected to dominate the SOC 
(with the possible exception of the CB). 
This is smaller than the typical inter-band energies for the 
$\textnormal{MoX}_2$ materials  and therefore the different bands are 
only weakly hybridized by the SOC. For the heavier $\textnormal{WX}_2$ compounds 
the VB spin-splitting is $425-460\, \textnormal{meV}$,  indicating 
that some matrix elements may not be small any more 
with respect to band-edge energy differences. One is therefore tempted 
to perform first a diagonalization of the SOC Hamiltonian (see Table \ref{tbl:SOC-at-K}), 
to obtain the eigenstates $|\Psi_{\mu,\mu'}^{b}, s \rangle$ which will be some linear
combination of the original basis states $|\Psi_{\mu}^{b}, s \rangle$, and then 
perform the $\mathbf{k}\cdot\mathbf{p}$ expansion and the perturbation 
calculation for the external electric field using this new basis.  
Diagonalization of the Hamiltonian (\ref{tbl:SOC-at-K}) is possible if one neglects 
the matrix elements $\Delta_{v-3,c+1}$, $\Delta_{v-3,c+2}$ and $\Delta_{v-2,c+2}$ 
between remote bands. The eigenstates are linear combinations of a symmetric and 
an antisymmetric basis vector.  However, the subsequent calculations in  
Secs.~\ref{subsec:kp-matrix} and \ref{subsec:efield} as well as the final L\"owdin partitioning 
are more tractable if we do not make this diagonalization and stay with the original basis
states throughout the calculations. The two approaches give the same results in the leading 
order of the ratio of the various SOC matrix elements and band-edge energy differences.  
For $\textnormal{MoX}_2$ compounds the approach outlined below is adequate,   
for the heavier  $\textnormal{WX}_2$ materials it still gives reasonable results, but the 
numerical estimates for, e.g., the effective $g$-factor might have to be revised, once 
experimental and theoretical consensus is reached regarding the magnitude of the 
band gap and SOC band splittings.  

\begin{widetext}
\begin{table}[htb]
\begin{tabular}{l|ccccccc}\hline\hline\vspace*{-0.8em}
 & & & & &  & & \\
$H_{\rm so}^{K}$  & $|\Psi_{A'}^{vb},s\rangle$ & $|\Psi_{E_1^{'}}^{cb},s\rangle$ & $|\Psi_{E_2^{'}}^{vb-3},s\rangle$ & 
       $|\Psi_{E_2^{'}}^{cb+2},s\rangle$ & $|\Psi_{E_1^{''}}^{vb-2},s\rangle $ & $|\Psi_{E_2^{''}}^{vb-1},s\rangle$ &
       $|\Psi_{A^{''}}^{cb+1},s\rangle$ \\
 \vspace*{-0.8em}
 & & & & & & & \\
 \hline \vspace*{-0.8em}
  & & & & & & &  \\
 \vspace*{-0.8em}
 $|\Psi_{A'}^{vb},s\rangle$ & $s_z\Delta_{v}$ & $0$ & $0$ & $0$ & $s_{-}\Delta_{v,v-2}^{}$ & 
$s_{+} \Delta_{v,v-1}^{}$ & $0$ \\
  & & & & & & &  \\       
\vspace*{-0.8em}
 $|\Psi_{E_1^{'}}^{cb},s\rangle$ & $0$ & $s_z\Delta_{c}^{}$ & $0$ & $0$ & $0$ & $s_{-}\Delta_{c,v-1}^{}$ & $s_{+}\Delta_{c,c+1}^{}$ \\
  & & & & & & &  \\         
\vspace*{-0.8em}     
$|\Psi_{E_2^{'}}^{vb-3},s\rangle$  & $0$  & $0$ & $s_z\Delta_{v-3}$ &  $s_{z}\Delta_{v-3,c+2}$ & $s_{+}\Delta_{v-3,v-2}$ & $0$ & 
$s_{-}\Delta_{v-3,c+1}$ \\
 & & & & & & &  \\         
\vspace*{-0.8em}     
 $|\Psi_{E_2^{'}}^{cb+2},s\rangle$ & $0$ & $0$ & $s_{z}\Delta_{v-3,c+2}^{*}$ & $s_z \Delta_{c+2}$ & $s_{+}\Delta_{c+2,v-2}^{}$ 
 & $0$ & $s_{-}\Delta_{c+2,c+1}^{}$\\
 & & & & & & &  \\         
\vspace*{-0.8em} 
 $|\Psi_{E_1^{''}}^{vb-2},s\rangle $& $s_{+}\Delta_{v,v-2}^{*}$ & $0$ & $s_{-}\Delta_{v-3,v-2}^{*}$ & $s_{-}\Delta_{c+2,v-2}^{*}$ 
 &  $s_z\Delta_{v-2}$ & $0$ & $0$ \\
 & & & & & & &  \\         
\vspace*{-0.8em} 
$|\Psi_{E_2^{''}}^{vb-1},s\rangle$ & $s_{-}\Delta_{v,v-1}^{*}$ & $s_{+}\Delta_{c,v-1}^{*}$ & $0$ & $0$ & $0$ & $s_z\Delta_{v-1}$ & $0$\\
& & & & & & &  \\         
\vspace*{-0.8em} 
 $|\Psi_{A^{''}}^{cb+1},s\rangle$ & $0$ & $s_{-}\Delta_{c,c+1}^{*}$ & $s_{+}\Delta_{v-3,c+1}^{*}$ & $s_{+}\Delta_{c+2,c+1}^{*}$ 
 & $0$ & $0$ & $s_z\Delta_{c+1}$\\
& & & & & & &  \\ 
\hline\hline
 \end{tabular}
 \caption{SOC matrix of TMDCs at the $K$ point in the seven-band model.}
 \label{tbl:SOC-at-K}
\end{table}
\end{widetext}

The SOC Hamiltonian at $K'$ can be obtained by making the following substitutions:
$\Delta_{b}\rightarrow \Delta_{b}^{*}$, 
$\Delta_{b,b'}\rightarrow  \Delta_{b,b'}^{*}$, $s_{\pm}\rightarrow - s_{\mp}$, $s_z\rightarrow -s_z $.
These relations follow from the fact the orbital wave functions at $K$ and $K'$ are connected by 
time-reversal symmetry, i.e.,
$|\Psi_{\mu}^{b}(K)\rangle = \hat{K}_0 |\Psi_{\mu'}^{b}(K')\rangle$, where $\hat{K}_0$ denotes 
complex conjugation. Consider, as an example, a matrix element 
$\langle\Psi_{\mu}^{b}(K')| \hat{L}_z |\Psi_{\mu'}^{b'}(K')\rangle$. 
\begin{eqnarray*}
 \langle  \Psi_{\mu}^{b}(K')| \hat{L}_z |\Psi_{\mu'}^{b'}(K')\rangle & = &
 \langle\hat{K}_0  \Psi_{\nu}^{b}(K)| \hat{L}_z |\hat{K}_0 \Psi_{\nu'}^{b'}(K)\rangle \\
 &=& \langle\hat{K}_0  \Psi_{\nu}^{b}(K)|  \hat{L}_z \hat{K}_0 \Psi_{\nu'}^{b'}(K)\rangle \\ 
 &=& \langle\hat{K}_0  \Psi_{\nu}^{b}(K)| (-1) \hat{K}_0 [ \hat{L}_z \Psi_{\nu'}^{b'}(K)]\rangle \\ 
 &=& -\langle [\hat{L}_z \Psi_{\nu'}^{b'}(K)]| \Psi_{\nu}^{b}(K)\rangle \\
 &=& - (\langle \Psi_{\nu}^{b}(K)| \hat{L}_z \Psi_{\nu'}^{b'}(K)\rangle)^{*}.
\end{eqnarray*}
Here we have made use of $\hat{K}_{0} \hat{L}_z=- \hat{L}_z \hat{K}_{0}$.
Relations for the matrix elements involving the operators $\hat{L}_{\pm}$ can be obtained 
by noting that $\hat{K}_{0} \hat{L}_{\pm} =- \hat{L}_{\mp} \hat{K}_{0}$ and therefore 
$\langle  \Psi_{\mu}^{b}(K')| \hat{L}_{\pm} |\Psi_{\mu'}^{b'}(K')\rangle = 
- (\langle\Psi_{\nu}^{b}(K)| \hat{L}_{\mp} |\Psi_{\nu'}^{b'}(K)\rangle)^{*}
$.

%##############################################################################################################

\subsection{$\mathbf{k}\cdot\mathbf{p}$ matrix elements at the $K$ ($K'$) points}
\label{subsec:kp-matrix}

The Hamiltonian
$ 
\mathcal{H}_{\mathbf{k}\cdot\mathbf{p}}=\frac{1}{2}\frac{\hbar}{m_e} (q_{+} \hat{p}_{-} + q_{-} \hat{p}_{+})
$
has  non-zero matrix elements only between states $|\Psi_{\mu}^{b}, s \rangle$ and  
$|\Psi_{\mu'}^{b'}, s \rangle$ which are either both
symmetric or antisymmetric with respect to the mirror operation $\sigma_h$. 
For the discussion in the main text we only need the matrix elements between symmetric states.
These matrix elements, which are diagonal in the spin-space, have  already been obtained in 
Ref.~\onlinecite{our-mos2}, but  for convenience they are replicated in Table \ref{tbl:kp-at-K}. 
We note that in addition to $\hat{p}_{\pm}$,  another operator due to SOC   appears  in the calculation 
of the $\mathbf{k}\cdot\mathbf{p}$ matrix elements\cite{dresselhaus-book,winkler-book}, but it can be neglected.  
The diagonal elements in  Table \ref{tbl:kp-at-K} are the band-edge energies. 

%\begin{widetext}
\begin{table}[htb]
\begin{tabular}{l|cccc}\hline\hline\vspace*{-0.8em}
 & & & & \\
$H_{\mathbf{k}\cdot\mathbf{p}}^{K}$  & $|\Psi_{A'}^{vb},s\rangle$ & $|\Psi_{E_1^{'}}^{cb},s\rangle$ & $|\Psi_{E_2^{'}}^{vb-3},s\rangle$ & 
       $|\Psi_{E_2^{'}}^{cb+2},s\rangle$  \\
 \vspace*{-0.8em}
 & & & &  \\
 \hline \vspace*{-0.8em}
  & & & &  \\
 \vspace*{-0.8em}
 $|\Psi_{A'}^{vb},s\rangle$ & $\vareps_v$ & $\gamma_3 q_{-}$ & $\gamma_2 q_{+}$ & $\gamma_4 q_{+}$ \\
  & & & &  \\       
\vspace*{-0.8em}
 $|\Psi_{E_1^{'}}^{cb},s\rangle$ & $\gamma_3^* q_{+}$ & $\vareps_c$ & $\gamma_5 q_{-}$ & $\gamma_6 q_{-}$  \\
  & & & & \\         
\vspace*{-0.8em}     
$|\Psi_{E_2^{'}}^{vb-3},s\rangle$  & $\gamma_2^{*} q_{-}$  & $\gamma_5^* q_{+}$ & $\vareps_{v-3}$ &  $0$  \\
 & & & &  \\         
\vspace*{-0.8em}     
 $|\Psi_{E_2^{'}}^{cb+2},s\rangle$ & $\gamma_4^{*} q_{-}$ & $\gamma_6^{*} q_{+}$ & $0$ & $\vareps_{c+2}$ \\
 & & & &  \\         
\hline\hline
 \end{tabular}
 \caption{The $\mathbf{k}\cdot\mathbf{p}$ matrix elements between symmetric states at the $K$ point.}
 \label{tbl:kp-at-K}
\end{table}
%\end{widetext}

The matrix elements at the $K'$ point can be obtained with the substitutions $\gamma_i \rightarrow \gamma_i^{*}$ and 
$q_{\pm}\rightarrow -q_{\mp}$. This follows from 

\begin{eqnarray*}
 \langle  \Psi_{\mu}^{b}(K')|  \mathcal{H}_{\mathbf{k}\cdot\mathbf{p}} |\Psi_{\mu'}^{b'}(K')\rangle  &=&  
 \langle\hat{K}_0  \Psi_{\nu}^{b}(K)|  \mathcal{H}_{\mathbf{k}\cdot\mathbf{p}} |\hat{K}_0 \Psi_{\nu'}^{b'}(K)\rangle\\
 &=&  \langle\hat{K}_0  \Psi_{\nu}^{b}(K)| (-1) \hat{K}_0 [ \mathcal{H}_{\mathbf{k}\cdot\mathbf{p}}  \Psi_{\nu'}^{b'}(K)]\rangle \\ 
 &=& - \langle\mathcal{H}_{\mathbf{k}\cdot\mathbf{p}} \Psi_{\nu'}^{b'}(K)]| \Psi_{\nu}^{b}(K)\rangle \\
 &=& - (\langle \Psi_{\nu}^{b}(K)|  \mathcal{H}_{\mathbf{k}\cdot\mathbf{p}} \Psi_{\nu'}^{b'}(K)\rangle)^{*}.
\end{eqnarray*}

As mentioned in Ref.~\onlinecite{our-mos2}, concrete values for the $\gamma_i$ parameters can be obtained either 
from fitting the band dispersion or using the Kohn-Sham orbitals to  evaluate directly the matrix 
elements $\langle \Psi_{\mu}^{b}|\hat{p}_{\pm}| \Psi_{\mu'}^{b'} \rangle$. The latter can be done, e.g.,
with the help of \textsc{castep} code (see Appendix \ref{sec:comp-detail} for computational details).
To estimate the effective valley and spin $g$-factor (Sec.~\ref{subsec:el-Ham-deriv}) and the Bychkov-Rashba SOC 
parameter (Sec.~\ref{subsec:BR-Ham-details}) we will need the value of $\gamma_3$, 
for which the two approaches give similar results.

\textbf{External magnetic field}
 
 The effects of  an external magnetic field in the  $\mathbf{k}\cdot\mathbf{p}$ formalism 
 can be obtained by using the Kohn-Luttinger prescription\cite{dresselhaus-book}, 
 which amounts to replacing the numbers $q_x$, $q_y$ in the above formulas
 with the operators $\hat{\mathbf{q}}=\frac{1}{i}\boldsymbol{\nabla}+\frac{e}{\hbar}\mathbf{A}$, 
 where $\mathbf{A}$ is the  vector potential and $e>0$ is the magnitude of the electron charge.
 Note that due to this replacement 
 $\hat{q}_+$ and  $\hat{q}_-$ become non-commuting operators and 
 their order has to be preserved  when one folds down the above  multi-band Hamiltonian 
 to obtain a low-energy effective Hamiltonian. Using the Landau gauge to describe a homogeneous,
 perpendicular magnetic field, the commutation relation is 
% \begin{equation}
 $
 [\hat{q}_-,\hat{q}_+] =\frac{2 e B_z }{\hbar}.
 $ 

%\end{equation}

%%%%%%%%%%%%%%%%%%%%%%%%%%%%%%%%%%%%%%%%%%%%%%%%%%%%%%%%%%%%%%%%%%%%%%%%%%%%%%%%%%%%%%%%%%%%%%%%%%%%%%%%%%%%%%%%%%%

\subsection{External electric field}
\label{subsec:efield}

In order to derive the Bychkov-Rashba SOC, we assume that 
a homogeneous, perpendicular external electric field is  present, which 
can be  described by the Hamiltonian  $U(z)=e E_z z$. 
It breaks the mirror symmetry $\sigma_h$ and therefore 
couples symmetric and antisymmetric basis states, while the matrix elements between states of the
same symmetry are zero. The full symmetry at the $K$ point is 
lowered from $C_{3 h}$ to $C_3$, i.e., the three-fold rotational symmetry is not broken. 
The  matrix elements of $H_{\rm U}^{K}$ between the symmetric and antisymmetric states 
are shown in Table \ref{tbl:Efield-at-K}. 
\begin{table}[htb]
\begin{tabular}{l|ccc}\hline\hline\vspace*{-0.8em}
 & & & \\
$H_{\rm U}^{K}$ & $|\Psi_{E_1^{''}}^{vb-2},s\rangle $ & $|\Psi_{E_2^{''}}^{vb-1},s\rangle$ & $|\Psi_{A^{''}}^{cb+1},s\rangle$ \\
 \vspace*{-0.8em}
 & & &  \\
 \hline \vspace*{-0.8em}
  & & &  \\
 \vspace*{-0.8em}
 $|\Psi_{A'}^{vb},s\rangle$ &  $0$ & $0$ & $\xi_{v,c+1}$ \\
  & & &  \\       
\vspace*{-0.8em}
 $|\Psi_{E_1^{'}}^{cb},s\rangle$ & $\xi_{c,v-2}$ & $0$ & $0$ \\
  & & &   \\         
\vspace*{-0.8em}     
$|\Psi_{E_2^{'}}^{vb-3},s\rangle$  & $0$ & $\xi_{v-3,v-1}$ & $0$ \\
 & & &   \\         
\vspace*{-0.8em}     
 $|\Psi_{E_2^{'}}^{cb+2},s\rangle$  & $0$ & $\xi_{c+2,v-1}$ & $0$\\
 & & &   \\         
\hline\hline
 \end{tabular}
 \caption{Matrix elements of the external electric field at the $K$ point between symmetric and antisymmetric states.}
 \label{tbl:Efield-at-K}
\end{table}

The matrix elements $\xi_{b,b'}=e E_z \langle \Psi_{\mu}^{b}| {z}| \Psi_{\mu'}^{b'} \rangle =  e E_z \zeta_{b,b'}$ 
are in general complex numbers. The magnitude of $\zeta_{b,b}$ can be calculated using the band-edge Kohn-Sham orbitals, 
 as in Ref.~\onlinecite{neil}, where this approach was used to estimate the 
electric-field-induced band gap in silicene (see Appendix \ref{sec:comp-detail} for computational details). 
Since the Kohn-Sham orbitals are defined only up to an arbitrary phase, from the actual 
calculations we cannot extract the real and imaginary parts of $\zeta_{b,b'}$. 
The matrix elements at the $K'$ point can be obtained by complex-conjugation of the $K$-point matrix elements.

%##################################################################################################################

\section{Effective low-energy Hamiltonian for the conduction band}
\label{sec:eff-Ham-details}

The total Hamiltonian of the system is then given by 
\begin{equation}
 \tilde{H}=\tilde{H}_{\mathbf{k}\cdot\mathbf{p}}+\tilde{H}_{\rm so}+\tilde{H}_{\rm U}.
\end{equation}
Since our seven-band model contains bands which are far from the CB, our next step is
to derive an effective Hamiltonian for the spin-split CB\@. This can be done by systematically
eliminating  all other bands using L\"owdin partitioning\cite{winkler-book}. 
Since the trigonal warping in the CB is weak, we consider terms up to second order in $\mathbf{q}$. 
We also keep the lowest  non-vanishing order in  the product of 
$\hat{q}_{\pm}$ and the SOC and electric field matrix elements.   

At the $K$ point one finds that the effective Hamiltonian is given by
\begin{subequations}
\begin{eqnarray}
 \tilde{H}_{\rm el}^{K, s}&=& \frac{\hbar^2 \hat{q}^2}{2 m_e}+
               \frac{|\gamma_3|^2}{\vareps_{c}^{K,s}-\vareps_v^{K,s}}\,\hat{q}_+ \hat{q}_- \nonumber\\
               &+&  \left[
 \frac{|\gamma_5|^2}{\vareps_{c}^{K,s}-\vareps_{v-3}^{K,s}}+
 \frac{|\gamma_6|^2}{\vareps_{c}^{K,s}-\vareps_{c+2}^{K,s}}
 \right] \hat{q}_- \hat{q}_+, \label{cb-el-K}\\
 \tilde{H}_{\rm so,intr}^{K,s}&=& s \Delta_c^{K}
  +
 \frac{|\Delta_{c,c+1}|^2}{\vareps_c^{K,\uparrow}-\vareps_{c+1}^{K,\downarrow}} s_{+}s_{-}\nonumber\\
 &+& 
 \frac{|\Delta_{c,v-1}|^2}{\vareps_c^{K,\downarrow}-\vareps_{v-1}^{K,\uparrow}}s_{-}s_{+}
 \label{cb-so-intr-K}\\
 \tilde{H}_{\rm U}^{K, s }&=&\frac{|\xi_{c,v-2}|^2}{\vareps_c^{K,s}-\vareps_{v-2}^{K,s}},
 \label{cb-U-K}\\
 \tilde{H}_{\rm BR}^{K}&=& 
 \left(
 \begin{array}{cc}
  0 & \lambda_{BR}^{*}\, \hat{q}_{-}\\
  \lambda_{BR}^{}\,\hat{q}_{+} & 0
 \end{array}
 \right),
 \label{cb-BR-K}
\end{eqnarray}
\label{eff-Ham-cond-K}
\end{subequations}
 
whereas at the $K'$ point:
\begin{subequations}
\begin{eqnarray}
 \tilde{H}_{\rm el}^{K', s}&=&\frac{\hbar^2 \hat{q}^2}{2 m_e}+
            \frac{|\gamma_3|^2}{\vareps_{c}^{K',s}-\vareps_v^{K',s}}\,\hat{q}_- \hat{q}_+ \nonumber\\
               &+&  \left[
 \frac{|\gamma_5|^2}{\vareps_{c}^{K',s}-\vareps_{v-3}^{K',s}}+
 \frac{|\gamma_6|^2}{\vareps_{c}^{K',s}-\vareps_{c+2}^{K',s}}
 \right] \hat{q}_+ \hat{q}_-, \label{cb-el-Kp}\\
 \tilde{H}_{\rm so,intr}^{K',s}&=& s \Delta_c^{K'}
 +
 \frac{|\Delta_{c,c+1}|^2}{\vareps_c^{K',\downarrow}-\vareps_{c+1}^{K',\uparrow}} s_{-}s_{+}\nonumber\\
 &+ &
 \frac{|\Delta_{c,v-1}|^2}{\vareps_c^{K',\uparrow}-\vareps_{v-1}^{K',\downarrow}} s_{+}s_{-}
 \label{cb-so-intr-Kp}\\
 \tilde{H}_{\rm U}^{K',s}&=&\frac{|\xi_{c,v-2}|^2}{\vareps_c^{K',s}-\vareps_{v-2}^{K',s}},
 \label{cb-U-Kp}\\
 \tilde{H}_{\rm BR}^{K'}&=&
 \left(
 \begin{array}{cc}
  0 & \lambda_{BR}^{*} \hat{q}_{-}\\
  \lambda_{BR} \hat{q}_{+} & 0
 \end{array}
 \right).
 \label{cb-BR-Kp}
\end{eqnarray}
\label{eff-Ham-cond-Kp}
\end{subequations} 
In the above formulas $m_e$ is the bare electron mass and 
we have used the notation $\vareps_{b}^{K (K'),s}=\vareps_{b} + s \, \Delta_{b}^{K,(K')}$,
where $s= \pm 1$ is the spin quantum number, $\Delta_{b}^{K,(K')}$ are  the diagonal SOC matrix 
elements from Sec.~\ref{subsec:SOC} at the $K,(K')$ point, and  $\vareps_{b}$ are the band-edge energies 
defined in Sec.~\ref{subsec:kp-matrix}, i.e., not taking into account the SOC\@. 
For convenience, in Eqs.~(\ref{cb-so-intr-K}) and (\ref{cb-so-intr-Kp})
we introduced the shorthand notation $\uparrow$ for $s=1$ and $\downarrow$ for $s=-1$.
Making use of the fact that the $K$ and $K'$ valleys are connected by time-reversal symmetry (see Sec.~\ref{subsec:SOC}), 
we can write $\Delta_{b}^{K, (K')}=\tau \Delta_b$, where $\tau=1 (-1)$ for $K$ ($K'$), and we can introduce the
notation $\vareps_{b}^{\tau,s}=\vareps_b+\tau\, s\, \Delta_b$.

The first term in Eqs.~(\ref{cb-el-K}), (\ref{cb-el-Kp}) is the free-electron contribution\cite{dresselhaus-book,winkler-book}.  
Regarding the other terms in Eqs.~(\ref{cb-el-K}) and (\ref{cb-el-Kp}) which contain $\hat{q}_{+}$ and $\hat{q}_{-}$, 
we did not assume that  they  
%$\hat{q}_{+}$ and $\hat{q}_{-}$ 
commute; see Appendix~\ref{subsec:el-Ham-deriv}.
Note that   $\tilde{H}_{\rm el}^{}$, $\tilde{H}_{\rm so,intr}^{}$ and $\tilde{H}_{\rm U}^{}$ are 
diagonal in spin space, but  the Bychkov-Rashba Hamiltonian $\tilde{H}_{\rm BR}$ introduces coupling between 
$\uparrow$ and $\downarrow$. We now briefly discuss each of the terms appearing in 
Eqs.~(\ref{eff-Ham-cond-K}) and (\ref{eff-Ham-cond-Kp}).

%%%%%%%%%%%%%%%%%%%%%%%%%%%%%%%%%%%%%%%%%%%%%%%%%%%%%%%%%%%%%%%%%%%%%%%%%%%%%%%%%%%%%%%%%%%%%%%%%%%%%%%%%%%%%%%%%%%%%%

\subsection{Electronic effective Hamiltonian $H_{\rm el}$}
\label{subsec:el-Ham-deriv}

In  the electronic Hamiltonian $ H_{\rm el}^{}$ we have taken into account the fact that in the 
presence of an external magnetic field the operators $\hat{q}_{+}$ and  $\hat{q}_{-}$ do not commute. 
To obtain Eq.~(\ref{kp-and-intr-so}) in the manuscript, one has to use the commutation relation 
$
[\hat{q}_-,\hat{q}_+] =\frac{2 e B_z }{\hbar}
$
and re-write $\frac{\hbar^2\hat{q}^2}{2 m_e}$ as 
$
\frac{\hbar^2\hat{q}_{+}\hat{q}_{-}}{2 m_e}+\frac{\hbar e B_z}{2 m_e}.
$
One finds 
\begin{eqnarray}
  \tilde{H}_{\rm el}^{K,s}&=&\frac{\hbar^2 \hat{q}_+ \hat{q}_-}{2 m_{\rm eff}^{\tau=1,s}}+
  \frac{\hbar e B_z}{m_{\rm eff}^{\tau=1,s}}\nonumber\\ &-&
  \left(\frac{1}{2 m_e}+
  \frac{2 |\tilde{\gamma_3}|^2}{\vareps_{c}^{\tau=1,s}-\vareps_v^{\tau=1,s}}\right)
  \hbar e  B_z 
  \label{H_el-magn-K}
\end{eqnarray}
in the $K$ valley and 
\begin{eqnarray}
 \tilde{H}_{\rm el}^{K',s}&=&\frac{\hbar^2 \hat{q}_+ \hat{q}_-}{2 m_{\rm eff}^{\tau=-1,s}}\nonumber\\
  &+&
  \left(\frac{1}{2 m_e}+
  \frac{2 |\tilde{\gamma_3}|^2}{\vareps_{c}^{\tau=-1,s}-\vareps_v^{\tau=-1,s}}\right)
  \hbar e  B_z
  \label{H_el-magn-Kp}
\end{eqnarray}
in the $K'$ valley. The effective mass $m_{\rm eff}^{\tau,s}$ is given by  
\begin{eqnarray}
 \frac{1}{2 m_{\rm eff}^{\tau,s}}&=&\frac{1}{2 m_e}+\frac{|\tilde{\gamma_3}|^2}{\vareps_c^{\tau,s}-\vareps_v^{\tau,s}}\nonumber\\
 &+&
 \frac{|\tilde{\gamma_5}|^2}{\vareps_c^{\tau,s}-\vareps_{v-3}^{\tau,s}}+ 
 \frac{|\tilde{\gamma_6}|^2}{\vareps_c^{\tau,s}-\vareps_{c+2}^{\tau,s}}.
\end{eqnarray}
In the above formulas $\tilde{\gamma_i}=\gamma_i/\hbar$. 
The inverse of the effective  mass $m_{\rm eff}^{\tau,s}$ can be then re-written 
in terms of $m^{0}_{\rm eff}$ and $\delta m_{\rm eff}$, as 
shown below Eq.~(\ref{kp-and-intr-so}).

The difference $\delta m_{\rm eff}$ in the effective masses comes  mainly from the 
spin-splitting $\Delta_v$ and $\Delta_{c+2}$ of the VB and CB+2, respectively, 
other diagonal SOC matrix elements being much smaller. 
We attribute the heavier effective mass at the $K$ point to the $\uparrow$ band. 
This assignment is  based  on the following. (i) From DFT calculations we 
know that both the VB and the CB+2 are composed mainly of $d_{x^2-y^2}$ and 
$d_{xy}$ orbitals.  Using group theoretical  considerations we take  
a VB Bloch wave function  $\sim d_{x^2-y^2}-i d_{xy}$,
whereas in the case of the CB+2 the Bloch wave function is $\sim d_{x^2-y^2}+i d_{xy}$.
(ii) Taking into account (i) we assume that 
$
\Delta_v=\langle \Psi_{A'}^{vb}(K)|H_{\rm so}^{\rm at}|\Psi_{A'}^{vb}(K)\rangle < 0
$
and 
$
\Delta_{c+2}=\langle \Psi_{E_1^{'}}^{vb+2}(K)|H_{\rm so}^{\rm at}|\Psi_{E_1^{'}}^{cb+2}(K)\rangle > 0.
$
Regarding (i), we note that since the states at the $K$ point are related to the  states 
at $K'$ by time reversal, our choice for the VB Bloch wave function is 
equivalent to other choices in the literature\cite{yao,cappelluti} up 
to a possible re-labeling of the valleys $K\leftrightarrow K'$. 
The sign of $\Delta_v$, as shown below, affects the sign of the effective spin $g$-factor, 
therefore it should be possible to deduce it experimentally. 
(From symmetry considerations\cite{our-mos2,song} and 
FP results\cite{cappelluti} we also know that there is a small X-$p$ orbital contribution to the 
VB and CB+2 as well, but in contrast to the CB, which is discussed in Sec.~\ref{subsec:int-so-Ham-deriv},  
this can be neglected in the case of the VB and CB+2 spin-splitting.)

The physical meaning of the  term 
$
[2 |\tilde{\gamma}_3|^2/(\vareps_c^{\tau,s}-\vareps_v^{\tau,s})] \hbar e B_z
$ 
appearing 
in Eqs.~(\ref{H_el-magn-K}) and (\ref{H_el-magn-Kp}) is probably more transparent if one expands it 
in powers of $(\Delta_{c}-\Delta_{v})/(\vareps_c-\vareps_v)$, where $E_{bg}=\vareps_c-\vareps_v$ is the band 
gap in the absence of  SOC\@. The zeroth-order term yields the valley-splitting Hamiltonian
$\tilde{H}_{vl}^{\tau}=-\tau \tilde{g}_{vl} \mu_{B} B_z$, with 
%$g_{vl}=1+4 m^{0}_{\rm eff}|\tilde{\gamma}_3|^2/E_{bg}$. 
\begin{equation}
 \tilde{g}_{vl}=1+4 m_{e}|\tilde{\gamma}_3|^2/E_{bg}.
 \label{eq:gvalley}
\end{equation}

The higher-order terms in the expansion determine how the coupling of the spin to the magnetic field 
is modified due to the  strong SOC in TMDCs. Keeping the first-order term only one arrives  
at the Hamiltonian 
$
\tilde{H}_{sp}^{s}= \frac{1}{2} \, g_{so}^{\perp}\, \mu_{B}\,  B_z
$
where $g_{so}^{}$ is an out-of-plane effective spin $g$ factor, 
\begin{equation}
g_{so}^{\perp} \approx \,
 8 m_e |\tilde{\gamma}_3|^2 \frac{\Delta_c^{}-\Delta_v^{}} {(E_{bg})^2}, 
\label{eq:gsp-factors}
\end{equation}
where $m_e$ is the bare electron mass. 
The value of $\Delta_c^{}$, i.e., the spin splitting coming from the X-$p$ orbitals in 
the CB  (see Sec.~\ref{subsec:int-so-Ham-deriv})
is not known; however, we can safely assume that it is negligible with respect to $\Delta_v^{}$. 
As explained  above,  we assume that $\Delta_v^{}< 0$, so we find  that  
$g_{so}^{\perp}\approx 8 m_e |\tilde{\gamma}_3|^2 |\Delta_v|/(E_{bg}^2)$. 
We note that in the case of bulk semiconductors 
a similar formula to Eq.~(\ref{eq:gsp-factors}) is called  Roth's formula\cite{roth}.

 The relevant parameters 
$\Delta_v$, $|\gamma_3|$, and $E_{bg}$ to calculate $g_{vl}$ and $g_{so}^{\perp}$ are shown 
in Table \ref{tbl:g-factor-params}. 
\begin{table}[ht]
 \begin{tabular}{|c|c|c|c|c|}\hline
     & $\textnormal{MoS}_2$ & $\textnormal{WS}_2$  &  $\textnormal{MoSe}_2$ & $\textnormal{WSe}_2$ \\
  \hline
  $|\gamma_3^{}|$ [eV/\AA] & $3.01$ & $3.86$ & $2.51$ & $3.32$ \\
\hline
  $ 2 |\Delta_v|$ [eV] & $ 0.146 $ & $0.42$ & $ 0.184 $ & $0.456$ \\
  \hline
  $ E_{bg}$ [eV] & $1.85$ & $1.98$ & $1.624$ & $1.736$ \\
  \hline
\end{tabular}
\caption{ Parameters appearing in the expressions for $g_{vl}$ and $g_{so}$ for different TMDCs.}
\label{tbl:g-factor-params}
\end{table}

The parameter  $\gamma_3$  %and $m_{\rm eff}^{0}$ 
was  obtained  with the help of Kohn-Sham orbitals (see Sec.~\ref{subsec:kp-matrix})
while the band gap $E_{bg}=\vareps_c-\vareps_v$ is readily available from our DFT calculations. 
We note that since  $E_{bg}$  is underestimated in DFT, 
the values of  $g_{vl}$ and $g_{so}$ shown in Table \ref{tbl:eff-g-factors} are overestimated.

%%%%%%%%%%%%%%%%%%%%%%%%%%%%%%%%%%%%%%%%%%%%%%%%%%%%%%%%%%%%%%%%%%%%%%%%%%%%%%%%%%%%%%%%%%%%%%%%%%%%%%%%%%%%%%%%%%%%%%%%%%%%

\subsection{Intrinsic SOC Hamiltonian $H_{\rm so,int}$}
\label{subsec:int-so-Ham-deriv}

Starting from Eqs.~(\ref{cb-so-intr-K}) and
(\ref{cb-so-intr-Kp}), it is easy to show that, apart from a constant term,
the intrinsic SOC Hamiltonian $H_{\rm so,int}^{}$ can be written as shown in Eq.~(\ref{kp-and-intr-so}), with 
$\Delta_{cb}=\Delta_c+(\omega_1-\omega_2)/2$, where $\omega_1 \approx |\Delta_{c,c+1}|^2/(\vareps_c-\vareps_{c+1})$ and
$\omega_2 \approx |\Delta_{c,v-1}|^2/(\vareps_c-\vareps_{v-1})$ and in the denominators we used 
$\vareps_b^{\tau,s}\approx\vareps_b$.

The spin-splitting in the CB has been discussed in  Refs.~\onlinecite{ochoa,our-mos2,gui-bin} before.  
Using  our  latest FP results, we revisit and expand our previous discussion\cite{our-mos2} of the problem. 
Generally, the intrinsic SOC Hamiltonian $H_{\rm so,int}$ has two contributions. 
One contribution  comes  from the  coupling of the CB to 
other, remote bands and is therefore second-order in the off-diagonal SOC matrix elements. 
In our seven-band model the couplings to  VB-1 and CB+1,
described by $\Delta_{c,c+1}$ and $\Delta_{c,v-1}$, 
are non-zero. These contributions are expected to be dominated by the metal 
$d$ orbitals. If one neglects the chalcogenide $p$ orbital admixing to the CB, 
these are the only terms that can explain the spin-splitting of the CB, which was 
found in FP calculations\cite{lambrecht,kosmider,our-mos2,gui-bin,zhu} and this 
was the motivation to consider  these second-order terms  in Ref.~\onlinecite{our-mos2}. 
For the $\uparrow$ states at the $K$ point the term $|\Delta_{c,c+1}|^2/(\vareps_c-\vareps_{c+1})$ 
 predicts a \emph{negative} shift. This would mean that the heavier $\uparrow$ band  
would be  lower in energy than the lighter $\downarrow$ band.  
In our DFT calculations this is indeed the case for $\textnormal{WS}_2$ and  $\textnormal{WSe}_2$, 
but not for $\textnormal{MoS}_2$ and  $\textnormal{MoSe}_2$.
However, from the orbital decomposition of the FP results (see, e.g., Ref.~\onlinecite{cappelluti}) 
we know that there is small chalcogenide $p$ orbital contribution to the CB as well. 
The X-$p$ orbitals, which have initially been neglected\cite{ochoa,our-mos2} in the discussion of 
the spin-splitting in the CB, give rise to the first term in Eqs.~(\ref{cb-so-intr-K}) and (\ref{cb-so-intr-Kp}) 
[the largest weight in  the CB  comes from the M-$d_{z^2}$ orbitals, but these carry 
no angular momentum, so they play no role in the SOC]\@. 
Taking  $\Delta_c^{}>0$ at the $K$ point (the corresponding Bloch wave function 
is an eigenfunction of $\hat{L}_z$ with positive eigenvalue: see Table IV in Ref.~\onlinecite{our-mos2}), 
the contribution of the X-$p$ orbitals to the  energy of the 
$\uparrow$ states is \emph{positive}. 
 Therefore a  plausible explanation of the presence/absence of the band crossing in the spin-split CB  
 for  $\textnormal{MoX}_2$/$\textnormal{WX}_2$ materials is that these two
 contributions compete. Namely, from Eqs.~(\ref{cb-so-intr-K}) and (\ref{cb-so-intr-Kp}) 
 it is  clear that the X-$p$ orbitals contribute to the spin splitting in first order, whereas remote bands 
contribute in second order; therefore it is not obvious which  is dominant.  
It is possible  that for $\textnormal{MoX}_2$ materials
the first, X-$p$ orbital related  term is larger, whereas in the case of $\textnormal{WX}_2$, 
which contains a heavier metal, the second term is larger, explaining the difference  
between the  $\textnormal{MoX}_2$ and $\textnormal{WX}_2$ materials 
regarding the energy of the heavier/lighter CB 
(this possibility has  recently been  also mentioned  in  Ref.~\onlinecite{gui-bin}).

In addition, the  X-$p$ orbital contribution to the CB 
spin-splitting seems  to be the simplest way to explain the difference
between the spin-splitting of  $\textnormal{MoS}_2$ and $\textnormal{MoSe}_2$:  
in our DFT calculations we find that it is larger in $\textnormal{MoSe}_2$ 
($\Delta_c^{\textnormal{MoSe}_2}\approx \,23\, \textnormal{meV}$) which contains a heavier chalcogenide than in 
$\textnormal{MoS}_2$ ($\Delta_c^{\textnormal{MoS}_2}\approx 3 \textnormal{meV}$). 
On the other hand, the  above reasoning would suggest that because of the competition between
the two SOC terms of different origins, the  splitting in $\textnormal{WS}_2$  
($\Delta_c^{\textnormal{WS}_2}\approx 38\textnormal{meV}$) 
should be larger than in  $\textnormal{WSe}_2$ ($\Delta_c^{WSe_2}\approx 46\textnormal{meV}$), 
which is not the case according to our DFT calculations. This might be related to the 
larger orbital weight of the M-$d$ orbitals in the relevant bands in the case of $\textnormal{WSe}_2$. 
In any case, the detailed understanding of the SOC in the CB  requires further study.

%%%%%%%%%%%%%%%%%%%%%%%%%%%%%%%%%%%%%%%%%%%%%%%%%%%%%%%%%%%%%%%%%%%%%%%%%%%%%%%%%%%%%%%%%%%%%%%%%%%%%%%%%%%%%%%%%%%

\subsection{Band-edge shift  $H_{\rm U}$}
\label{subsec:U-Ham-deriv}

The Hamiltonian $H_U$ in Eqs.~(\ref{cb-U-K}) and (\ref{cb-U-Kp}),  
describes the dependence of the band edge on the external electric field. 
An order-of-magnitude estimate can be obtained by calculating $\zeta_{c,v-2}$ using
LDA Kohn-Sham orbitals, generated by the \textsc{castep} code.
As one can see from Table~\ref{tbl:H_U-nums}, it is a small effect for the electric field 
values ($E_z\lesssim 10^{-2}$ V/\AA), where the perturbation theory should be valid,
and therefore we neglect it. 
We note, that as one can see in  Eqs.~(\ref{cb-U-K}) and (\ref{cb-U-Kp}), the value of $H_{\rm U}$ 
also depends (indirectly) on $E_{bg}$. 
The band gap, according to $GW$ calculations\cite{lambrecht,ashwin-2,komsa,yakobson,li_yang}, 
is most likely to be underestimated by our DFT-LDA 
calculations. On the other hand, $\xi_{c,v-2}$ is probably  overestimated, because 
screening is neglected in our perturbative Kohn-Sham-orbital-based calculations. As a consequence,
the values shown in Table \ref{tbl:H_U-nums}  overestimate  the real value of $H_{\rm U}$. 
This conclusion is supported by our preliminary DFT results on the $E_z$ dependence of $E_{bg}$ obtained 
by the \textsc{castep} code.
\begin{table}[ht]
 \begin{tabular}{|c|c|c|c|c|}\hline
    & $\textnormal{MoS}_2$ & $\textnormal{WS}_2$  &  $\textnormal{MoSe}_2$ & $\textnormal{WSe}_2$ \\
  \hline
  $H_{\rm U}$[meV]  & $24.6\, E_z^2$ & $2.4\, E_z^2$ & $30.3\, E_z^2$ & $3.0\, E_z^2$ \\
  \hline
\end{tabular}
\caption{Band-edge shift  $H_{\rm U}$ in meV, if $E_z$ is expressed in V/\AA. 
 }
\label{tbl:H_U-nums}
\end{table}

The shift of the band edge is, in principle, spin-dependent,
but as one can see it from Eqs.~(\ref{cb-U-K}) and (\ref{cb-U-Kp}), 
this is  a higher-order effect and  can be safely neglected.

%%%%%%%%%%%%%%%%%%%%%%%%%%%%%%%%%%%%%%%%%%%%%%%%%%%%%%%%%%%%%%%%%%%%%%%%%%%%%%%%%%%%%%%%%%%%%%%%%%%%%%%%%%%%%%%%%%%

\subsection{Bychkov-Rashba Hamiltonian  $H_{\rm BR }$}
\label{subsec:BR-Ham-details}

Finally, we discuss the Bychkov-Rashba Hamiltonian [ Eqs.~(\ref{cb-BR-K}) and (\ref{cb-BR-Kp})]. 
It is a sum of several terms, each having the same structure and related to the 
matrix elements $\xi_{v,c+1}$, $\xi_{v-3,v-1}$, $\xi_{c+1,v-1}$ and $\xi_{c,v-2}$.  
Using L\"owdin-partitioning, one finds for the most important term at the $K$ point, 
\begin{widetext}
\begin{subequations}
\begin{eqnarray}
 \tilde{H}_{\rm BR}^{(1),K}&\approx&%\frac{1}{(\vareps_c+\Delta_c^S-(\vareps_v-\Delta_v))(\vareps_c+\Delta_c^S-(\vareps_{c+1}-\Delta_{c+1}))}
 \frac{1}{(\vareps_c^{}-\vareps_v^{\downarrow})(\vareps_c^{}-\vareps_{c+1}^{})}
% \frac{1}{\vareps_{cv}^{\uparrow,\downarrow}\,\delta\vareps_{c,c+1}^{\uparrow,\downarrow}}
 \left(\gamma_3^*q_{+}\xi_{v,c+1} s_{-}\Delta_{c,c+1}^{*}+\gamma_3 q_{-}\xi_{v,c+1}^{*} s_{+} \Delta_{c,c+1}\right)
 \nonumber\\
 &=&  (\lambda_{\rm BR}^{(1),r}+ i \lambda_{\rm BR}^{(1),i})\, q_{+} s_{-} +  
      (\lambda_{\rm BR}^{(1),r}- i \lambda_{\rm BR}^{(1),i})\,  q_{-} s_{+} \nonumber\\
  &=& \lambda_{\rm BR}^{(1),r} \left(s_x q_x+s_y q_y\right) + \lambda_{\rm BR}^{(1),i}\left(s_y q_x -s_x q_y\right)
  \label{eq-szunyogh}\\
 &=&
  \left( \begin{array}{cc}      
        0 & (\lambda_{\rm BR}^{(1)})^{*} \,q_- \\
        \lambda_{\rm BR}^{(1)} \, q_+ & 0
       \end{array}
       \right).
       \label{eq-operators}
\end{eqnarray}
\end{subequations}
\end{widetext}
To make the results more transparent, in the above formula we have neglected the 
spin-splittings of the CB and  CB+1, which are much smaller than the splitting
of the VB. 
The product $\gamma_3^* \xi_{v,c+1} \Delta_{c,c+1}^{*}$ is in general a complex number and 
therefore the Bychkov-Rashba coupling constant 
\begin{equation}
 \lambda_{BR}^{(1)} = \frac{\gamma_3^* \xi_{v,c+1} \Delta_{c,c+1}^{*}}
 {(\vareps_c^{}-\vareps_v^{\downarrow})(\vareps_c^{}-\vareps_{c+1}^{})}
 \label{lambda1}
\end{equation}
is also complex. By separating the real and imaginary part of $\lambda_{BR}^{(1)}$ one can write 
$H_{\rm BR}^{(1),K}$ in the more familiar form shown in Eq.~(\ref{eq-szunyogh}).

One can estimate the magnitude of  $\lambda_{BR}^{(1)}$ in  the following way.  
As mentioned in Sec.~\ref{subsec:efield}, one can calculate the magnitude of $\zeta_{v,c+1}^{z}$ 
and the parameter $\gamma_3$
using the band-edge Kohn-Sham  orbitals (see Table \ref{tbl:lambda1-params}).  
The band-edge energies $\vareps_{c}^{\uparrow,\downarrow}$, 
$\vareps_{v}^{\downarrow}$ and  $\vareps_{c+1}^{\uparrow,\downarrow}$
are known from  DFT-LDA band structure calculations; we have collected their values 
in Table \ref{tbl:lambda1-params}. 
Unfortunately, the off-diagonal SOC matrix element $\Delta_{c,c+1}$ %, $\Delta_{c,v-1}$ 
is not directly given by the DFT calculations. However, information about the 
weight of the M-$d$ orbitals in each of the bands can be obtained from DFT 
computations and therefore we can relate this matrix 
element to $\Delta_v$, because the dominant contribution to  the SOC should  come from the M-$d$ orbitals. 
Since the M-$d$ orbital weight  in both the CB and the  CB+1 band is similar to the  one in the VB,  
we  take $|\Delta_{c,c+1}|\lesssim |\Delta_v|$. 

\begin{table}[ht]
 \begin{tabular}{|c|c|c|c|c|}\hline
     & $\textnormal{MoS}_2$ & $\textnormal{WS}_2$  &  $\textnormal{MoSe}_2$ & $\textnormal{WSe}_2$ \\
   \hline
  $|\xi_{v,c+1}|$ [eV\AA] & $0.54 \,E_z$ & $ 0.6\, E_z$ & $ 0.57\, E_z $ & $0.64\, E_z$ \\
  \hline
  $\vareps_c^{}-\vareps_v^{\downarrow}$ [eV] & $1.77$ & $1.71$ & $ 1.54 $ & $1.44$ \\
  \hline
  $\vareps_c^{}-\vareps_{c+1}^{}$ [eV] & $-1.16$ & $-1.33$ & $-0.925$ & $-1.14$ \\
  \hline
\end{tabular}
\caption{ Parameters appearing in Eq.~(\ref{lambda1}) for different TMDCs.  
$\xi_{v,c+1}$ was 
calculated using DFT-LDA Kohn-Sham orbitals; the other parameters are obtained from DFT-LDA band 
structure calculations. $E_z$ is in units of V/\AA.}
\label{tbl:lambda1-params}
\end{table}

Similar procedure can be  performed to estimate the terms proportional to the other non-zero $\xi_{b,b'}$  
matrix elements as well.
We have found that the magnitude of  these further  terms are 
significantly smaller than that of $\lambda_{BR}^{(1)}$, mainly because of the pre-factors which are 
inversely proportional to the product of band-edge energy differences between remote bands. 
Therefore, as an order-of-magnitude estimate of the strength of the Bychkov-Rashba SOC,  one 
can just use $\lambda_{BR}^{(1)}$.  Taking  values for $|\gamma_3|$ from Table \ref{tbl:g-factor-params} 
and for the other parameters from Table \ref{tbl:lambda1-params},  
one finally arrives at the results shown in Table \ref{tbl:lambda_br-values}.

The method outlined  here  most likely overestimates the real values of the Bychkov-Rashba parameters.
In addition to the uncertainties in the values of the SOC matrix elements and the $\gamma_i$ parameters, 
there are two other sources of error: i) the calculation of $\zeta_{b,b'}^z$  did  not take into account 
screening effects (see  Ref.~\onlinecite{neil})  and 
ii) according to $GW$ calculations, the real band gap is larger then the DFT one, and this affects 
the energy denominators in the above formulas.

%%%%%%%%%%%%%%%%%%%%%%%%%%%%%%%%%%%%%%%%%%%%%%%%%%%%%%%%%%%%%%%%%%%%%%%%%%%%%%%%%%%%%%%%%%%%%%%%%%%%%%%%%%%%%%%%%%%

\section{Eigenfunctions of the $\alpha_{-}$ and  $\alpha_{+}$ operators}
\label{sec:alpha-eigenfunc}

Considering the functions 
$
g_{a,l}(\rho,\varphi)=e^{i l \varphi} \rho^{\frac{|l|}{2}} e^{-\frac{\rho}{2}} M(a,|l|+1,\rho)
$
one can show that 
\begin{equation}
 \hat{\alpha}_{-}\,g_{a,l}(\rho,\varphi) = 
 \left\{
 \begin{array}{cc}
  \frac{a}{|l|+1}\, g_{a+1,l-1}(\rho,\varphi)  & l\le 0, \\
    l\, g_{a,l-1}(\rho,\varphi)  &  l>0,
 \end{array}
 \right.
\end{equation}
and 
\begin{equation}
 \hat{\alpha}_{+}\,g_{a,l}(\rho,\varphi) = 
 \left\{
 \begin{array}{cc}
    l \, g_{a-1,l+1}(\rho,\varphi)  & l <  0, \\
    \left(1-\frac{a}{m+1}\right)\, g_{a,l+1}(\rho,\varphi)  &  l \ge 0.
 \end{array}
 \right.
\end{equation}

To prove these relations, one may use the following identities for the confluent hypergeometric
functions: 
\begin{eqnarray}
 \partial_{\rho}M(a,b,\rho)&=&\frac{a}{b}\,M(a+1,b+1,\rho)\\
 (b-a)\,M(a,b+1,\rho)&=&b\,M(a,b,\rho)\nonumber\\
                   &-&b \,\partial_{\rho}M(a,b,\rho),\\
 (b-1)\,M(a,b-1,\rho)&=& (b-1)\,M(a,b,\rho)\nonumber\\
		     &+&\rho\,\partial_{\rho}M(a,b,\rho),\\
 (b-1)\,M(a-1,b-1,\rho)&=&(b-1-\rho)\,M(a,b,\rho)\nonumber\\
                     &+&\rho\,\partial_{\rho}M(a,b,\rho).
\end{eqnarray}

%%%%%%%%%%%%%%%%%%%%%%%%%%%%%%%%%%%%%%%%%%%%%%%%%%%%%%%%%%%%%%%%%%%%%%%%%%%%%%%%%%%%%%%%%%%%%%%%%%%%%%%%%%%%%%%%%%%

\section{Computational details}
\label{sec:comp-detail}

The band structure calculations were performed with the \textsc{vasp} code\cite{vasp} using 
the LDA\@.
%local density approximation of DFT. 
The plane-wave cutoff energy was 600 eV. We used a $12 \times 12$ 
Monkhorst-Pack $\mathbf{k}$-point grid in the 2D plane to relax the geomety and a 
$24 \times 24$ grid to calculate the band structure.
The artificial periodicity in the vertical direction was $20$\,\AA. 
The optimized lattice parameter $a_0$ for each TMDC is shown in Table \ref{tbl:lattice-values}.
\begin{table}[htb]
 \begin{tabular}{|c|c|c|c|c|}\hline
     & $\textnormal{MoS}_2$ & $\textnormal{WS}_2$  &  $\textnormal{MoSe}_2$ & $\textnormal{WSe}_2$ \\
  \hline
  $a_0$ [\AA] & $3.129$ & $3.131$ & $3.253$ & $3.253$ \\
\hline
\end{tabular}
\caption{DFT-LDA lattice parameters.}
\label{tbl:lattice-values}
\end{table}
 
The matrix elements of the momentum operator $\hat{p}_{\pm}$ and  the Hamiltonian describing the 
perpendicular electric field were evaluated  within the LDA  using the \textsc{castep}  
code\cite{castep}, because 
 the necessary plane-wave coefficients of the Kohn-Sham orbitals  at the band edges
 were readily accessible in the output of \textsc{castep}. 
We used norm-conserving pseudopotentials, a plane-wave cutoff energy of 2177\,eV, %80 Ha
 an artificial periodicity of  %30 Bohr 
 15.9\,\AA\   in the vertical direction and a $21 \times 21$ 
Monkhorst-Pack mesh. The optimized lattice parameters were similar 
to those found  in the \textsc{vasp} calculations.

%%%%%%%%%%%%%%%%%%%%%%%%%%%%%%%%%%%%%%%%%%%%%%%%%%%%%%%%%%%%%%%%%%%%%%%%%%%%%%%%%%%%%%%%%%%%%%%%%%%%%%%%%%%%%%%%%%%%

%\bibliography{tmdc-qdot-biblio}

\begin{thebibliography}{75}
\bibitem [{eff()}]{effmass-fit}%
  \BibitemOpen
  \href@noop {} {}\bibinfo {note} {The fitting was performed in a range around
  the $K$ point which corresponds to $\approx 6\%$ of the $\Gamma-K$
  distance.}\BibitemShut {Stop}%
\end{thebibliography}

%

\end{document}